\documentclass[useAMS,usenatbib]{mn2e}
\usepackage{amsmath}
\usepackage{graphicx}
\usepackage{xfrac}
\usepackage{caption}
\usepackage{array}
\usepackage{longtable}
\usepackage{pdflscape}
\usepackage{subcaption}
\title[The stellar mass of Brightest Cluster Galaxies]{The evolution in the stellar mass of Brightest Cluster Galaxies over the past 10 billion years}
\author[Bellstedt et al.]{Sabine Bellstedt$^{1,2}$\thanks{Email: sbellstedt@swin.edu.au}, Chris Lidman$^{1,3}$, Adam Muzzin$^{4}$, Marijn Franx$^{5}$, \newauthor Susanna Guatelli$^{1}$,  Allison R. Hill$^{5}$, Henk Hoekstra$^{5}$, Noah Kurinsky$^{6,7}$, Ivo Labbe$^{5}$, \newauthor Danilo Marchesini$^{6}$, Z. Cemile Marsan$^{6}$, Mitra Safavi-Naeini$^{1}$, Crist\'{o}bal Sif\'{o}n$^{5}$, \newauthor Mauro Stefanon$^{5}$, Jesse van de Sande$^{8}$, Pieter van Dokkum$^{9}$ and Catherine Weigel$^{6}$\\
$^{1}$School of Physics, University of Wollongong, Wollongong, NSW 2522, Australia\\
$^{2}$Centre for Astrophysics and Supercomputing, Swinburne University of Technology, Hawthorn VIC 3122, Australia\\
$^{3}$Australian Astronomical Observatory, North Ryde, NSW, Australia\\
$^{4}$Kavli Institute for Cosmology, University of Cambridge, Madingley Road, Cambridge, CB3 0HA, United Kingdom\\
$^{5}$Leiden Observatory, Leiden University, NL-2300 RA Leiden, Netherlands\\
$^{6}$Department of Physics and Astronomy, Tufts University, Medford, MA 02155, USA\\
$^{7}$Kavli Institute for Particle Astrophysics and Cosmology, Stanford University, 450 Serra Mall, Stanford, CA 94305, USA\\
$^{8}$Sydney Institute for Astronomy, School of Physics, University of Sydney, NSW 2006, Australia\\
$^{9}$Department of Astronomy, Yale University, New Haven, CT 06520, USA\\}
\begin{document}

\date{Accepted 2016 May 16. Received 2016 May 15; in original form 2015 December 1}

\pagerange{\pageref{firstpage}--\pageref{lastpage}} \pubyear{2015}

\maketitle

\label{firstpage}

\begin{abstract}

Using a sample of 98 galaxy clusters recently imaged in the near
infra-red with the ESO NTT, WIYN and WHT telescopes, supplemented with
33 clusters from
the ESO archive, we measure
how the stellar mass of the most massive galaxies in the universe, namely
Brightest Cluster Galaxies (BCG), increases with time. Most
of the BCGs in this new sample lie in the redshift range
$0.2<z<0.6$, which has been noted in recent works
to mark an epoch over which the growth in the stellar mass of BCGs
stalls.  From this sample of 132 clusters, we create a subsample of 102
systems that includes only those clusters that have estimates of the
cluster mass. We combine the BCGs in this subsample with 
BCGs from the literature, and find that the growth in stellar
mass of BCGs from 10 billion years ago to the present epoch
  is broadly consistent with recent semi-analytic and semi-empirical models.
As in other recent studies, tentative evidence indicates that the stellar mass growth rate of BCGs may be slowing in the past 3.5 billion years. 
Further work in
  collecting larger samples, and in better comparing observations with theory using mock images
  is required if a more detailed comparison between the models and the data is to be made.

\end{abstract}

\begin{keywords}
galaxies: elliptical and lenticular, cD - galaxies: evolution - galaxies: clusters: general

\end{keywords}

\section{Introduction}

Brightest Cluster Galaxies (BCGs) are the brightest and most massive
galaxies in the universe. They form within galaxy clusters, and
generally lie near the bottom of the cluster gravitational potential
well. They have unique properties, including extended light profiles,
and they are brighter than the cluster luminosity function leads us to
expect \citep{Loh06, vonderLinden07, Shen13}. These properties
differentiate them from other elliptical galaxies.

Most BCGs can be readily identified in observations as a result of their
brightness and dominance within a galaxy cluster. Additionally, N-body
simulations can be carried out to create mock galaxy clusters, which
also contain readily-identifiable BCGs. These observed and simulated
BCGs can be directly compared, and this allows us to test models that
describe the growth of these BCGs - a task that is difficult to do
with galaxies in general as a result of the large variety of different
types of galaxies, with varying formation histories.

Initially, there was considerable disagreement between
  the models and the observations, with models predicting a factor of
  three increase in the stellar mass of BCGs between $z=1$ and today
  \citep[hereafter referred to as DLB07]{DeLucia07}, and observations
  showing little growth over the same redshift interval
  \citep[see][for example]{Stott10}.  While more recent models
  \citep{Tonini12, Shankar14, Shankar15} and observations \citep{Lin13,Lidman12}
  are now in better agreement with one another with both predicting or showing a
  doubling of the stellar mass since $z\sim1$, there is still some
  disagreement as to when this growth occurs. In the semi-empirical
  model of \citet{Shankar15}, the stellar mass of BCGs continues to
  increase to the present day. However, in the semi-analytic model of
  \citet{Tonini12}, the growth appears to stall after $z\sim
  0.4$. There is some observational support for the second
  model. \citet{Lin13} find that most of of the growth since
  $z\sim1.5$ occurs in the redshift range $0.5<z<1.5$. Similarly,
  \citet{Oliva14} find no significant growth in the range
  $0.09<z<0.27$, and \citet{Inagaki14}, who explores the redshifts
  range $0.2<z<0.4$, find an increase of between 2 and 14\%. In
  contrast to these results, \citet{Bai14} find an increase of 50\%
  between $z=0.5$ and $z=0.1$, and \citet{Zhang15} find an increase of 35\% between 
  $z=1$ and the present day.

The aim of this study is to make a more detailed measurement
of the growth of BCGs using new data that covers the
redshift interval over which the growth appears to stall.
The paper is outlined as follows. \S \ref{sec:Data}
describes the data used and the steps used to process them, and \S
\ref{sec:StelMasDet} outlines how the stellar masses of the BCGs and
clusters within the sample were determined. The analysis of the data
is carried out in \S \ref{sec:Analysis}, and the discussion and
conclusions are presented in \S \ref{sec:Discussion} and \S
\ref{sec:Conclusion} respectively. Throughout this paper we use Vega
magnitudes, and assume a $\Lambda$CDM cosmology with $\Omega_{M}=0.3$,
$\Omega_{\Lambda}=0.7$ and $H_0=70
\text{\ km\ s}^{-1}\text{\ Mpc}^{-1}$.

\section{Data}
\label{sec:Data}

We utilise a sample of 98 newly imaged galaxy clusters from the
RELICS\footnote{REd Lens Infrared Cluster Survey} survey within this
study. The data were collected during 6 observing runs on three
instruments over a period spanning from October 2013 to March
2015. The instruments utilised were the SofI\footnote{Son of ISAAC}
camera on the New Technology Telescope at the ESO La Silla Observatory
in Chile \citep{Moorwood98}, WHIRC\footnote{WIYN High-Resolution
  Infrared Camera} on the WIYN telescope at the Kitt Peak National
Observatory \citep{Miexner10} and LIRIS\footnote{Long-slit
  Intermediate Resolution Infrared Spectrograph} on the William
Herschel Telescope in La Palma, Spain. The observing runs are
summarised in Table \ref{ObservingRuns}.

\begin{table}
	\centering
	\caption{Observing runs}
	\label{ObservingRuns}
	\begin{tabular}{@{}llr}
		\hline
		\hline
		Instrument & Telescope & Dates \\
		\hline
		WHIRC & WIYN & 2013 October 11-14 \\
		LIRIS & WHT &2014 December 12-14 \\
		SofI & NTT & 2014 January 18-21 \\
		LIRIS & WHT & 2014 October 3-4 \\
		SofI & NTT & 2014 December 5-8 \\
		LIRIS & WHT  &  2015 March 6-8 \\
		\hline
	\end{tabular}
\end{table}

RELICS  uses   massive  clusters  from  the   SPT\footnote{South  Pole
  Telescope}   \citep{Carlstrom11},   ACT\footnote{Atacama   Cosmology
  Telescope}  \citep{Swetz11},  MACS\footnote{MAssive Cluster  Survey}
\citep{Ebeling01} and  C1G \citep{Buddendiek15} cluster  catalogues as
gravitational telescopes to search for lensed
compact early-type galaxies (also know as red nuggets). Many of the
clusters within the RELICS survey are drawn from the ``Weighting the
Giants" Survey \citep{vonderLinden14}, which has provided deep Subaru
imaging of a selection of MACS clusters, in addition to a sample of
Abell clusters.  Clusters in the MACS sample \citep{Ebeling10} are
X-ray selected, whereas the clusters in the SPT/ACT
samples \citep{Staniszewski09, Williamson11, Marriage11,
  Hasselfield13} have been discovered through the Sunyaev-Zel'Dovich
effect.  Clusters in the C1G cluster catalogue were
  selected from a joint search of ROSAT all sky survey and
  data-release 8 of the Sloan Digital Sky Survey. RELICS clusters are
massive, and have a median mass of $10^{15}M_{\odot}$.

This sample was then augmented by carrying out a search of the ESO
archives of the SofI instrument. These data were obtained from
September 1998 to February 2012, and this search resulted in an
additional 31 clusters being added to the sample.

The clusters are all imaged in the Ks-band, with typical exposure
times of $\sim 2700$s, which result in 5-$\sigma$ depths
  of 19.5\,mag. These observations are all deeper than what is
necessary for the analysis of BCGs, as they were designed to detect
background galaxies gravitationally lensed by the cluster. At the
redshift range of the sample, BCGs have Ks-band magnitudes of 12-17
mag, and therefore each BCG is detected with a minimum signal-to-noise
ratio of 50.

\begin{table*}
	\caption[Observational Summary]{Observational summary \\(see https://github.com/SabineBellstedt/Bellstedt2016---Table-2-Observational-Summary for full table)\\ Redshift Sources:\\ (1) \citet{Mantz10} (2) \citet{Mann12}  (3) \citet{Menanteau13} (4) \citet{Kristian78} (5) \citet{Gioia98} (6) \citet{Ebeling07}  (7) \citet{Abell89}  (8) SDSS, DR8  (9) \citet{Menanteau10}  (10) \citet{Werner07}    (11) \citet{Jones09}  (12) NASA Astrophysical Database  (13) \citet{vanWeeren12} (14) \citet{Williamson11}  (15) \citet{Story11}  (16) \citet{Aghanim12}  (17) \citet{Sifon13}  (18) \citet{Ebeling10} (19) \citet{Applegate14} (20) \citet{Ruel14} (21) \citet{Buddendiek15} (22) \citet{Wen13} (23) \citet{Piffaretti11} (24) \citet{Vanderlinde10} (25) \citet{vonderLinden14} (26) \citet{Sifon13} (27) \citet{Brodwin10}}
	\label{ObservationalSummary}
	\begin{tabular}{@{}lccccccc}
		\hline
		\hline
		Cluster & RA & Dec & $z_{spec}$ & $z_{phot}$ & z Source & Instrument/Telescope & Exposure Time \\
		 & J2000 & J2000 &  &  &  & & [s] \\
		\hline
SPT-CL-J0000-5748 & 00:01:00.04 & -57:48:20.7 & 0.702 & ... & (20) & NTT/SofI & 2700 \\
MACS-J0011.7-1523 & 00:11:42.80 & -15:23:18.34 & 0.379 & ... & (18) & WHT/LIRIS & 2700 \\
MACS-J0014.3-3022 & 00:14:15.82 & -30:22:14.4 & 0.308 & ... & (1) & SofI/NTT & 12000 \\
MACS-J0014.3-3022 & 00:14:17.26 & -30:22:34.8 & 0.308 & ... & (1) & SofI/NTT & 12000 \\
ACT-CL-J0014.9-0057 & 00:14:54.00 & +00:57:10.1 & 0.533 & ... & (7) & NTT/SofI & 2700 \\
		\hline
	\end{tabular}
\end{table*}

\begin{table*}
	\caption{Instrument summary}
	\label{InstrumentSummary}
	\begin{tabular}{@{}llccc}
		\hline
		\hline
		Instrument & Telescope & Pixel Scale & FoV & Detector \\
		& & ["] & ['] & \\
		\hline
		SofI & NTT & 0.288 & 4.9   & $1024\times1024$ Rockwell Hawaii HgCdTe\\
		WHIRC & WIYN & 0.099 & 3.7  & $2048\times2048$ Raytheon Virgo HgCdTe\\
		LIRIS & WHT & 0.251 & 4.3 & $1024\times1204$ Hawaii HgCdTe \\
		\hline
	\end{tabular}
\end{table*}

\subsection{Data Reduction}

The procedures used in the reduction of the data are standard, and
largely follow the steps as outlined by \citet{Lidman08}. Briefly, the
pedestal in the images was removed using dark frames, pixels were
normalised using dome flats and the sky was removed using a moving
median stack of the science data, using our own python scripts and
tasks in IRAF\footnote{IRAF is distributed by the National Optical
  Astronomy Observatory, which is operated by the Association of
  Universities for Research in Astronomy (AURA) under a cooperative
  agreement with the National Science Foundation.}.

Zero points were determined by using stars from the 2MASS point source
catalogue \citep{Skrutskie06}. Typically, between 4 and 30 stars were
present in each image, and only unsaturated stars with high quality
measurements were selected to measure the zero point.

For data taken with LIRIS, we first had to unscramble the image pixels
in the FITS
header\footnote{http://www.ing.iac.es/astronomy/instruments/liris/detector.html}. There
was also a residual shade pattern in the sky subtracted images. This
was removed by subtracting the median of the data along detector
rows. A similar technique was used to remove the crosstalk from bright
stars in SofI images.

\subsection{Data Analysis}

To estimate the magnitudes of galaxies in
each cluster, we run
{\sc SExtractor} \citep{Bertin96} on each image, and use {\tt
  MAG\_AUTO} as a measure of the magnitude. {\tt MAG\_AUTO} is a
Kron-like magnitude \citep{Kron80} with an elliptical aperture.

Since the galaxy clusters within the sample were selected to be
massive, most of the clusters have
a large BCG. 
The BCG in each cluster was identified visually as the
largest, brightest galaxy in the cluster. BCGs are typically in the
centre with extended galaxy haloes, unlike their surrounding
galaxies. We verified the visual BCG selection by ensuring that these
were the galaxies which were indeed the brightest as measured by {\tt
  MAG\_AUTO}. In some clusters, there appear to be two BCGs of comparable brightness,
such as the cluster MACS J0014.3-3022. In such cases, both BCGs were included in the sample.

There are a small number of cases in which the BCGs were not clearly identifiable, usually because 
of foreground contamination, these clusters are excluded from 
analysis. The excluded clusters are Abell 521, ACT-CL J0018.2-0022, 
ACT-CL J0228.5+0030, ACT-CL J0301.1-0110, ACT-CL J0250.1+0008, MACS 
J2243.3-0935, RX J1132+00 and SPT-CL J0615-5746.

\subsection{Additional Samples}

To augment the sample used throughout this paper, we include the
sample used by \citet[hereafter L12]{Lidman12}. This sample includes 103 BCGs presented by
\citet{Stott08} over the redshift range $0.04\leq z \leq 0.83$, 20
BCGs by \citet{Stott10} over the higher redshift range of $0.81 \leq z
\leq 1.46$ and 5 BCGs from \citet{Collins09} over the redshift range
$1.22 \leq z \leq 1.46$. The study by L12 produced two new BCG samples
referred to as the CNOC1 \citep{Yee96} and SpARCS samples. The SpARCS
sample includes 12 BCGs spanning the redshift interval $0.867\leq z
\leq 1.630$, and the CNOC1 sample has 15 BCGs over the lower redshift
range $0.193\leq z \leq 0.547$.

\begin{figure}
	\centering
	\includegraphics[width=70mm]{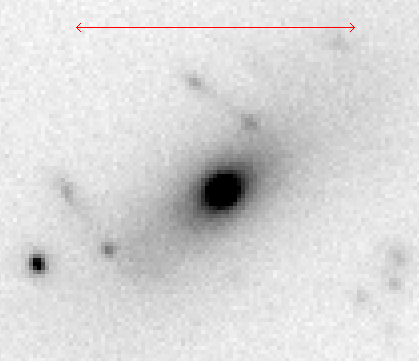}
		\caption{
                  A SofI image in the Ks band of the BCG
                  in the cluster Abell 1553. North is pointing up and east
                  to the left, and the arrow has a length of 20",
                  which at the cluster redshift of $z=0.165$
                  corresponds to a distance of 57 kpc. The extended
                  light profile of this BCG is clearly seen in many of
                  the BCGs in our sample. Of particular note
                  in this BCG are the lens-like features offset from
                  the centre of the BCG.}
		\label{fig:ClusterExample}
\end{figure}

The photometric errors determined by {\sc SExtractor} tend to
underestimate the true error as correlated signal in the pixels is not
taken into account.  
Some of the clusters imaged in
the new set of data had previously been analysed by
L12. These clusters are;
Abell 1204, Abell 1553, Abell 1835, Abell
2390, Abell 68, MACS J0025.4-1222 and MACS J0454.10-0300. We use these
clusters to provide a more reliable
measure of the errors. The median
of the absolute difference in Ks-band magnitude of the BCGs in these
images is 0.22\,mag, which we have used as the magnitude error of the
BCGs in our sample.

\subsection{Systematic drifts in the photometry}
\label{sec:photometry}

Throughout this paper we use {\tt MAG\_AUTO} from SExtractor to
estimate magnitudes, which are then used to derive masses. 
In essence, it is an aperture magnitude, so by definition, it does not
measure the total magnitude of a galaxy. The amount of flux missed
depends on the intrinsic light profile of the galaxy and the
seeing. On simulated galaxies, L12 finds that {\tt MAG\_AUTO} misses
between 18\% and 35\% of the flux. Our results are not affected if the
amount of flux missed is independent of redshift. However, this
assumption may not be true, as the profile of BCGs and therefore the
amount of flux lost may change with redshift.

\begin{table}
	\centering
	\caption[GALFIT results]{Testing the difference between photometric magnitudes produced by {\tt MAG\_AUTO} and GALFIT to eliminate the presence of a measurement bias.}
	\begin{tabular}{ c  c  c  c  }
	\hline
	\hline
	Redshift &  Number of   & {\tt MAG\_AUTO} - & Scatter \\
	range &  clusters  & GALFIT  &  \\
	\hline
	0.00-0.25 & 10 & 0.52  & 0.37 \\
	0.25-0.40  & 5  & 0.44  & 0.17 \\
	0.40-0.80 & 10 &  0.57 &  0.33 \\
	0.00-0.80 & 25 & 0.54  & 0.33 \\
	\hline
	\end{tabular}
	\label{Table:GALFITResults}
\end{table}

In order to see if this is a large effect, we tried an alternative
approach in computing the magnitude of the BCGs in our
sample. Following L12, we ran version 3.0.4 of GALFIT on a subsample
of BCGs. For this test, we used the BCGs that were observed with SofI
and we split the BCGs into three redshift bins. The bin boundaries are
the same as those used later in the analysis,
 i.e.~$0 < z <
0.25$, $0.25 < z < 0.4$, and $0.4 < z < 0.8$. We then compared GALFIT
magnitudes with {\tt MAG\_AUTO}. The difference between the two is
large, with the GALFIT value being on average 0.54 mag brighter than the {\tt MAG\_AUTO} 
magnitude with a scatter of 0.33 mag when using 
all 25 BGCs. This is very similar to the difference reported in L12,
although our sample is a factor of three larger. When split into the
three redshift bins, there is no significant change in the average
difference between the three redshift bins. 

While this test is not definitive (as any systematic drift may affect
{\tt MAG\_AUTO} and GALFIT equally), it is suggestive that the 
 deviations from the models identified in Section
  \ref{sec:BCGGrowth} do not result from systematic uncertainties in the 
photometry.

\section{Determining Stellar Masses}
\label{sec:StelMasDet}

\subsection{BCG Stellar Masses}

\begin{figure*}
	\includegraphics[trim=20mm 0mm 20mm 0mm, width=110mm]{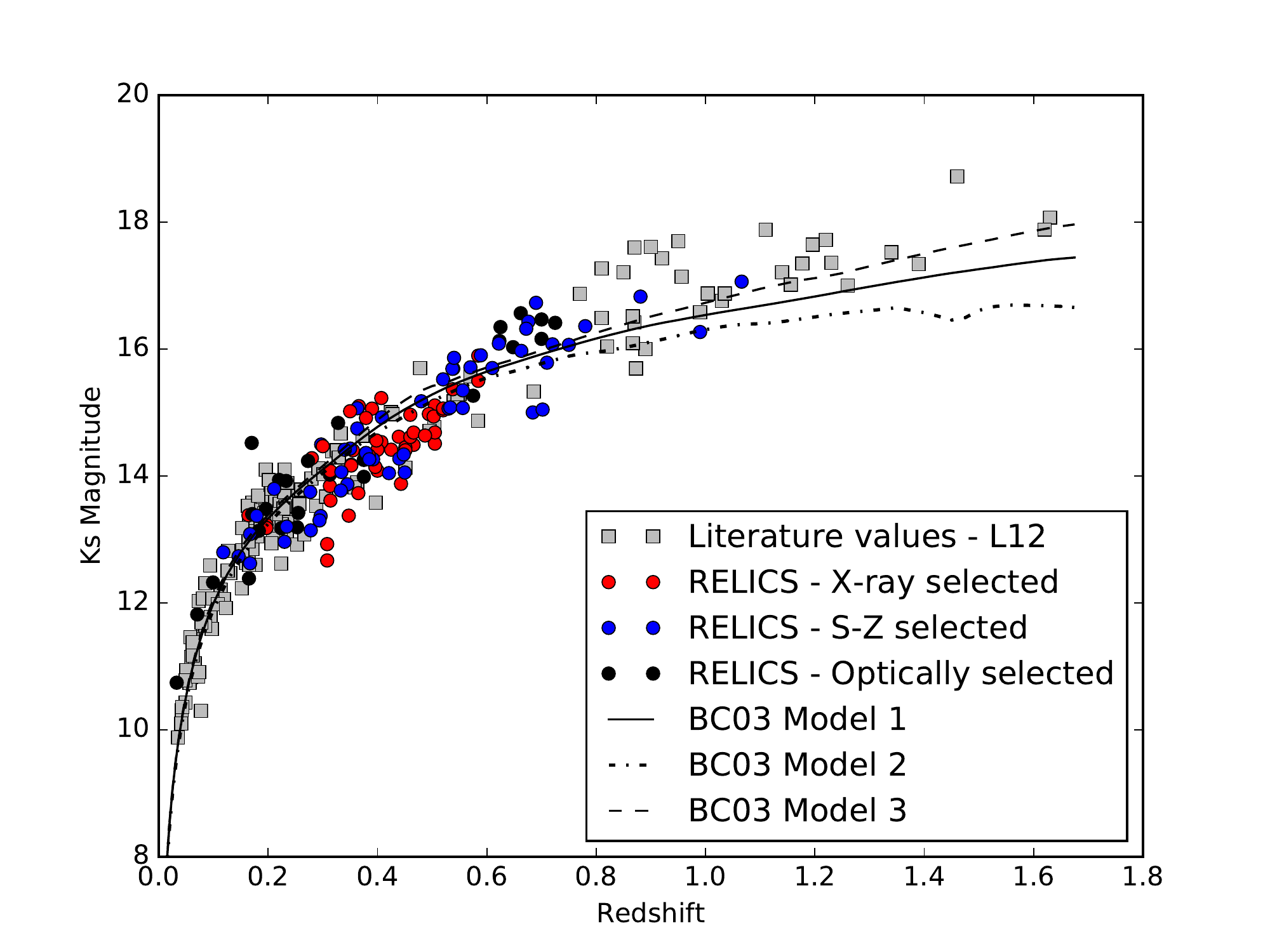}
		\caption{The observer-frame Ks-band magnitude of BCGs as a function of redshift. The data from this paper have been colour coded based on the selection method of the host cluster: X-ray selected clusters have been coloured in red, Sunyaev-Zel'dovich effect selected clusters have been coloured in blue, and the rest have been coloured in black. The stellar population model indicated as BC03 Model 1 is that from \citet{Lidman12}, with $z_f = 5$, $\tau=0.9$ and a metallicity split 60/40 between solar and 2.5 times solar, as outlined in the text. Other BC03 single burst models are indicated on the plot: BC03 Model 2 has  $z_f = 2$, and BC03 Model 3 has  $z_f = 5$.  These models have been normalised so that they agree with the low redshift sample. Note how most of the BCGs above $z\sim1$ lie above the model. }
		\label{fig:HubbleDiagram}
\end{figure*}

The stellar masses of BCGs were determined using the method developed
in L12, in which the observed Ks-band magnitude is converted into a
mass using stellar population synthesis models. In choosing the model,
L12 used the J-Ks colour of BCGs over the redshift range $0<z<1.5$ as
a constraint. From their comparison, it was determined that the
best-fitting stellar population model was a \citet{Bruzual03} model
with a Chabrier IMF \citep{Chabrier03}, a formation redshift of
$z_f=5$, a star formation rate e-folding time of 0.9 Gyr and a
composite metallically that is split 60/40 between solar and a
metallicity that is two and half times solar. Since a large portion of
the sample of this paper is in common with the sample of L12, we use
this model in our analysis. In converting from luminosity to mass, we
assume that the mass-to-light ratios of the BCGs are independent of
stellar mass.

Since $\sim6\%$ of the clusters within the sample do not yet have
spectroscopic
redshifts, the corresponding masses for the 
BCGs could not be calculated. Therefore, these clusters were not
considered further in this paper.

The observed Ks-band magnitudes of the full sample are plotted against
redshift in Figure \ref{fig:HubbleDiagram}. The new sample of this
study is shown as
red, blue and black circles, and can be seen to augment the
previously sparsely populated region of the plot in the redshift range
$0.4<z<0.8$.

In addition to the model described above, we plot two other models to display the effect of modifying the model. Both models, labelled as Model 2 and Model 3 in Figure \ref{fig:HubbleDiagram} are single burst models. Model 2 has a formation redshift of $z=2$ and has solar metallicity. Model 3 has a formation redshifts of $z=5$ and has metallicity that is two and half times solar. Since neither model is capable of describing the observed J-Ks colour (see Figure 2 in  L12), we do not consider these models further in this paper.

There are some preliminary observations that can be made from
Figure \ref{fig:HubbleDiagram}. The main one is that the BCGs at
$z>0.8$ (made up of the samples by \citet{Stott10}, \citet{Collins09}
and the SpARCS sample) tend to lie above the stellar population model.
Within the redshift interval $0.6<z<0.8$,
this trend ends. For the new sample of this paper, the opposite effect 
is noticed. While the effect is not so strong as for the high-z BCGs, 
the new BCGs are more likely to fall below the model than above, 
indicating that they are more massive than the model suggests. 
What is important to note however, is that larger clusters tend to host larger BCGs, 
and therefore it is not possible to make judgements about the growth of BGCs 
without accounting for their corresponding cluster masses. This is discussed further in the 
next section. 
Another observation to be noted is that X-ray selected clusters tend to host brighter BCGs. As will become clear in the next section, this occurs because the X-ray selected clusters in our sample are, on average, more massive than clusters selected by other means.

\subsection{Cluster Halo Masses}

Before the magnitude offsets seen in Figure
\ref{fig:HubbleDiagram} can be interpreted as an indication of
BCG stellar mass growth, it is important to note that the clusters
observed in our sample are particularly massive. As larger clusters
generally tend to host larger BCGs \citep[for example][]{Edge91, Burke00,
Brough08, Whiley08, Stott12} it is unsurprising that the BCGs of
this sample are brighter than the stellar population model in Figure
\ref{fig:HubbleDiagram}. In order to reduce the likelihood of
sustaining a large systematic error in our final calculation of BCG
growth, it is important to account for the masses of the host clusters.

Cluster masses for the new sample were calculated using a number of
methods. For clusters with published estimates of the X-ray luminosity
\citep{Henry92, Ebeling98, Menanteau10, Mantz10, Piffaretti11,
  Mann12, Menanteau13}, we applied the $L_X-M$
relation from \citet{Vikhlinin09} to calculate the corresponding
$M_{500}$ mass\footnote{$M_{\Delta}$ is defined as the mass measured
  in a region within which the average density is $\Delta$ times
  the critical density of the universe $\rho_c(z)$.}.

\begin{equation}
\begin{split}
	\ln L_X &= (47.392\pm 0.085)+(1.61\pm 0.14)\ln M_{500}  \\
	&+(1.850\pm 0.42)\ln E(z)-0.39\ln \left(\frac{h}{0.72}\right) \\
	&\pm (0.396\pm 0.039) 
	\label{eqn:L-M}
\end{split}
\end{equation}

If an X-ray temperature is
available for the cluster
\citep{Ebeling07, Mahdavi13}, then we apply the $T_X-M$ scaling relation from
\citet{Mantz10} to calculate the corresponding
$M_{500}$ mass.

\begin{equation}
\begin{split}
	 \log_{10}\left(\frac{kT}{keV}\right) &=(0.88\pm0.03) + \\
	 (0.49\pm0.04)\log_{10}&\left(\frac{E(z)M_{500}}{10^{15} M_{\odot}}\right) \pm (0.056\pm0.008) 
	\label{eqn:T-M}
\end{split}
\end{equation}

In both  equations \ref{eqn:L-M} and \ref{eqn:T-M},  $E(z)$ represents
the normalised  Hubble parameter,  given by  $E(z) =  H(z)/H_0$.  
For some clusters, estimates of $M_{500}$ were already available. In such cases, we use the 
these masses, and note the mass proxy that was used to compute these masses in 
Column 10 of Table
\ref{BCGClusterResults}.

In order to judge whether the proxy used can have a systematic effect on the calculated cluster masses,
we make comparisons for the calculated cluster masses for all clusters for which we have information
from multiple proxies. 
We measure the mean ratios between cluster masses measured using different methods, and note that these ratios are each consistent with one (see Table \ref{tab:MassComparison}). Using these ratios, we rescale each mass to be consistent with the mass calculated with X-ray luminosities. We have run our analysis with these scaled masses in addition to the original masses, and find no difference in our final results. We are therefore confident that we have not introduced any additional biases as a result of using multiple mass proxies listed in the literature.

\begin{table}
	\caption{Mean ratios between cluster masses measured using different methods.}
	\label{tab:MassComparison}
	\begin{tabular}{@{}ccc}
		\hline
		\hline
		Mass Proxy 1 & Mass Proxy 2 & Mean Ratio \\
		 &  & Proxy 1 / Proxy 2 \\
		\hline
		S-Z parameter & X-ray luminosity & $0.82\pm0.48$ \\
		X-ray temperature & X-ray luminosity & $1.25\pm0.53$ \\
		Gas mass & X-ray luminosity & $1.43\pm0.45$ \\
		Gas mass & X-ray temperature & $1.16\pm0.44$ \\
		\hline
	\end{tabular}
\end{table}

We then convert from $M_{500}$ to $M_{200}$ by assuming that the
cluster mass profile follows a Navarro-Frenk-White profile
\citep{Navarro97}, and then apply the mass conversion method outlined
in Appendix C of \citet{Hu03}.
Throughout this conversion, we assume a constant concentration parameter of $c=5$. We have carried out our analysis with altered concentration
parameter values, including a lower value of $c=4$ and a relation with mass and redshift \citep{Duffy08}:
\begin{equation}
c = 11.93  \left(\frac{M}{2\times10^{12}}\right)^{-0.09}(1+ z)^{-0.99}
\end{equation}
Since the effect of this on the cluster masses of our sample was negligible, we use $c=5$ throughout the study. 

Of the clusters that
make up the new sample, cluster mass data were
only available for 106 of the clusters. Since the consideration of the 
influence of the host cluster mass is imperative to the final
calculation of BCG growth, the remaining $\sim30$ clusters for which
cluster mass data could not be calculated were omitted from all
further calculations.

\begin{table}
	\caption{Data summary}
	\label{DataSummary}
	\begin{tabular}{@{}lc}
		\hline
		\hline
		Property & Number available \\
		& in new sample \\
		\hline
		Clusters imaged & 132 \\
		Clusters with redshifts & 124 \\
		Clusters with two BCGs & 11 \\
		Clusters with reliable mass calculations & 102 \\
		\hline
	\end{tabular}
\end{table}

Four clusters from \citet{Menanteau13} have
published X-ray luminosity
uncertainties that exceed 100\%,
and we therefore did not include these clusters
in our analysis. The excluded
clusters were ACT-CL-J0218.2-0041, ACT-CL-J2050.5-0055,
ACT-CL-J2302.5+0002 and ACT-CL-J0219+0022. This leaves us
with a sample of 102 clusters for which we have reliable mass
measurements.

\begin{figure}
	\includegraphics[trim=8mm 5mm 8mm 0mm, width=84mm]{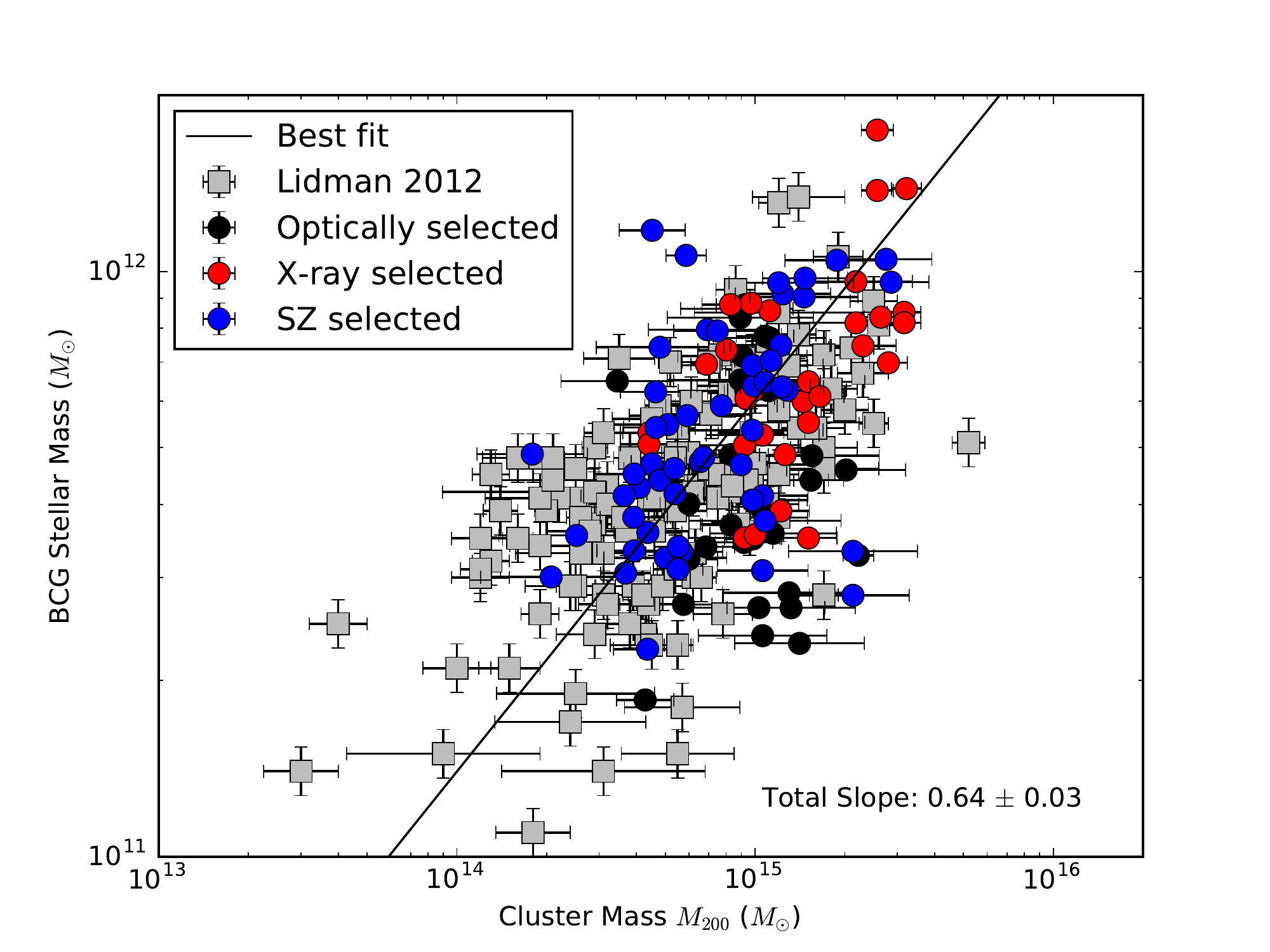}
		\caption{The correlation between the mass of the BCG,
                  and the mass of the cluster at the time at which it
                  was observed. Note that all cluster masses are
                  measured as $M_{200}$. }
		\label{fig:ClusterBCGRelation}
\end{figure}

The correlation between BCG stellar mass and the mass of the
host cluster is shown in Figure \ref{fig:ClusterBCGRelation}. The
positive correlation between the two variables is quite clear - a
power law fit to the relation of the form ${M}_{\bf BCG} = \beta
{M}_{\bf Cluster}^{\alpha}$ results in a best fit index of $\alpha =
0.64\pm0.03$, similar to that found by \citet{Lidman12}.
For the BCGs from the RELICS sample, we have coloured the points based on the selection method of the host clusters. Here one sees that the X-ray selected clusters tend to be larger, and therefore also host larger BCGs, while SZ selected clusters are smaller, with smaller BCGs. Interestingly, the optically selected clusters seem to host smaller BCGs for the given cluster mass. 

\section{Analysis}
\label{sec:Analysis}

Our method of computing the stellar mass growth of BCGs involves
comparing the median mass of BCGs in low- and high-redshift samples.
The positive correlation between BCG and cluster mass, as shown in Figure
\ref{fig:ClusterBCGRelation}, highlights the need to account for
cluster mass when making a measurement of the mass differences of BCGs 
in different redshift intervals. 

\subsection{Accounting for Cluster Masses}

\begin{figure*}
	\includegraphics[trim=15mm 5mm 15mm 3mm, width=160mm]{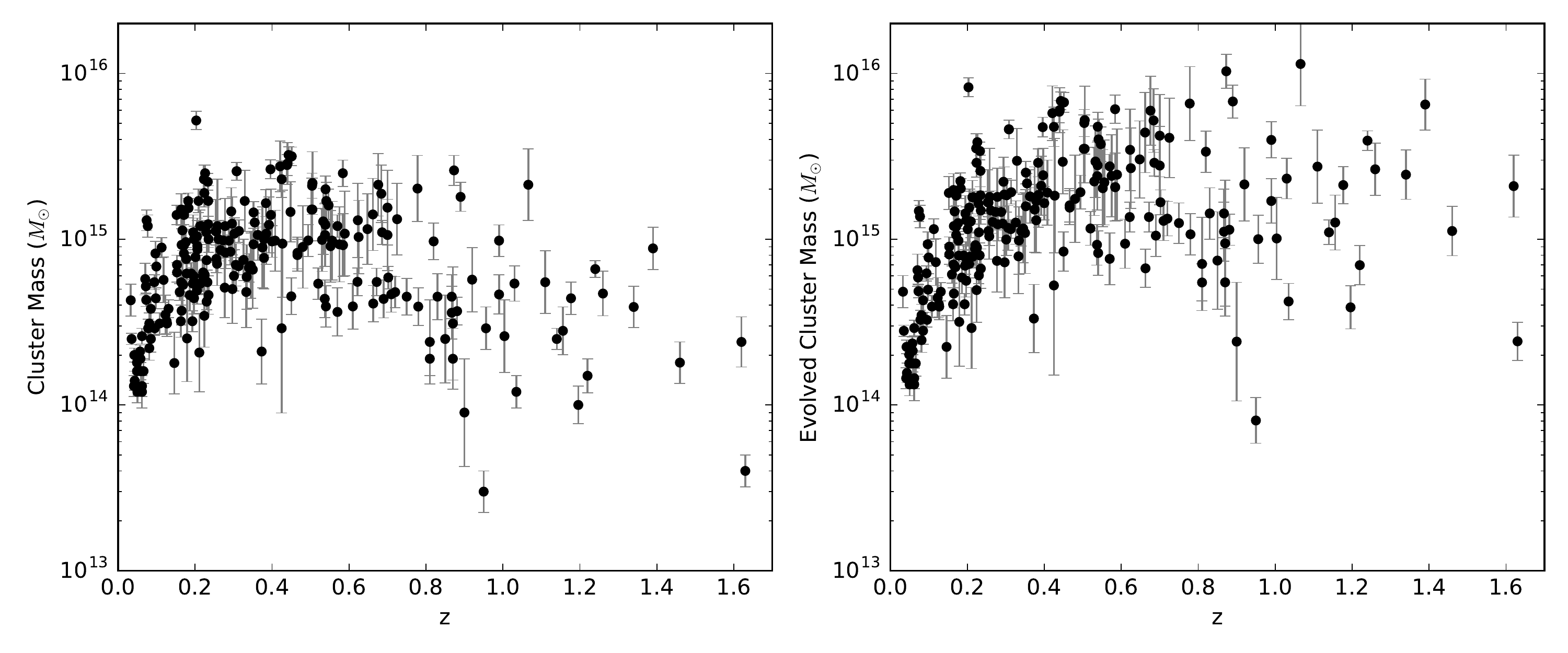}
		\caption{The distribution of cluster masses as a
                  function of redshift. Left: Masses at the epoch the
                  clusters were observed. Right: Masses by the present
                  day ($z=0$). A number of selection effects are
                  visible in these plots. The most massive clusters
                  are under-represented at lower redshifts. Similarly,
                  the least massive clusters are under-represented at
                  high redshifts. }
		\label{fig:EvolvedClusterMasses}
\end{figure*}

To make an unbiased comparison of the stellar mass of BCGs over 
different redshift intervals, we follow the procedure outlined in L12. 
To guarantee that BCGs originating from like-sized clusters are 
compared with each other, we ensure that during any comparison between 
BCGs in different redshift bins, the two samples have matching cluster 
mass distributions. 
Because clusters observed at high redshift will grow in mass
over time, it is necessary that the 
samples are matched based on their evolved cluster masses, rather than 
simply those measured at the cluster redshift. 
To achieve this, L12 
first computed the mass each cluster would have by today using the 
fitting formulae in \citet{Fakhouri10}. The effect this 
has on the masses is evident from the comparison of the 
left- and right-hand panels in Figure \ref{fig:EvolvedClusterMasses}. 

After defining two redshift intervals over which to measure the growth 
of the BCGs, two matched samples are produced by randomly selecting 
clusters from each of the two intervals without replacement until the 
cluster mass histograms of the two samples match. The median BCG mass 
in each sample is then compared. The error in the mass ratio is estimated by repeating 
the resampling 100 times.

The approach only compares one redshift interval  with another interval and not all
intervals simultaneously. Unfortunately, there are not enough clusters in the sample to 
match the cluster mass distributions in all four redshift intervals 
simultaneously, so we are forced to compare them in pairs.
We therefore select one redshift interval as a reference bin (marked in 
black in Figures \ref{fig:MassEvolutionSingle} and
\ref{fig:MassEvolution}), and calculate the growth relative to this point 
for each of the other bins individually.

Inevitably, some clusters will be rejected if they are over-represented in one 
redshift interval compared to the other.  The bin size selected to 
match these histograms needs to be small enough to match the overall 
cluster mass distributions, but large enough so that the rejection of 
clusters was not unnecessarily large. We applied a bin size of $4\times10^{14} M_{\odot}$.

\subsection{BCG Growth Calculation}
\label{sec:BCGGrowth}

Four redshift intervals are considered in our analysis:
$0.00<z<0.25$, $0.25<z<0.40$, $0.40<z<0.80$ and $0.80<z<1.60$.
We do two comparisons: One using the $0.00<z<0.25$ redshift bin 
as the comparison point, and another using the $0.25<z<0.40$ redshift bin as the comparison point. This allows
  us to scale the results at two separate redshifts for easier comparison with models.
We present the results of this analysis
in Table \ref{Table:GrowthResults}, and in Figures
\ref{fig:MassEvolutionSingle} and \ref{fig:MassEvolution}.

\subsubsection{The first scaling}

\begin{figure*}
	\includegraphics[trim={20mm, 0mm, 20mm, 0mm}, width=180mm]{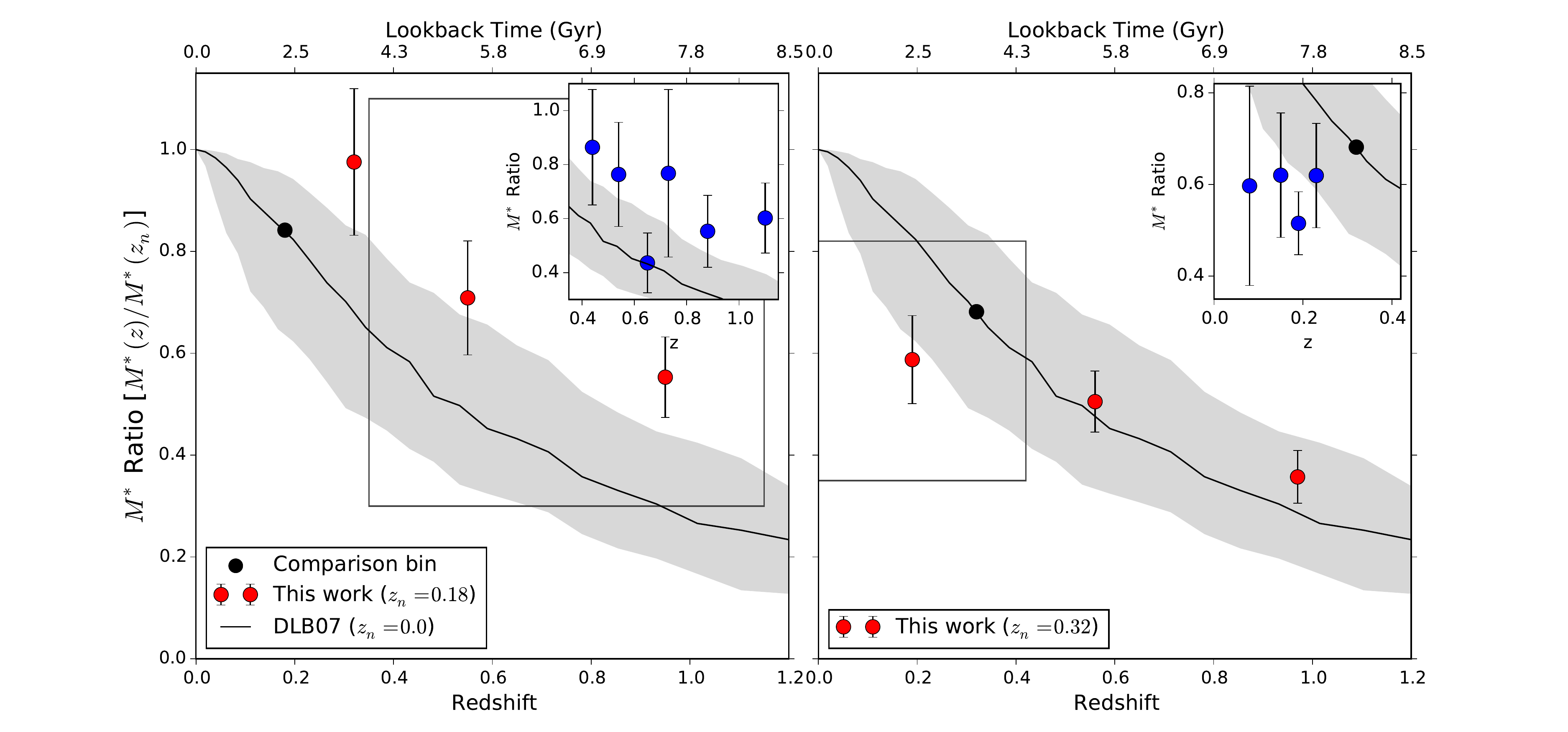}
		\caption{The evolution in BCG stellar mass as a
                  function of redshift normalised to one at $z=0$.
                  The DLB07 model is shown as the solid line and our
                  results are shown as the red, blue and black circles. 
		Red points represent the ratio between the stellar mass of BCGs in these 
		bins and the stellar mass of BCGs in the reference bin (black circles). 
		The left plot shows the results for the first scaling, using a reference bin with mean redshift $z=0.18$, 
		whereas the right plot shows the results for the second scaling with a reference bin with mean redshift $z=0.32$. 
		 The grey area shows the error range of the model to which our data has been compared. 
		The points of this work have been rescaled so that the black point lies on the DLB07 model. 
		It is apparent that within each of the scalings, the data disagrees with the model at different redshift 
		intervals. In the low-z scaling, the data above $z=0.2$ all sits above the model. Once the data is scaled to
		the mid-z interval however, the data above $z=0.2$ agree with the model, however the low-z range lies below	
		the model. 
		As a check that this trend does not simply result from our binning selection, we re-run our analysis for smaller bins
		in both the high-z and low-z region, and display these results as the blue points in the insets of the left and right plots
		respectively. Although the errors are larger, it is clear that the trends are the same. 
		}
		\label{fig:MassEvolutionSingle}
\end{figure*}

\begin{figure*}
	\includegraphics[trim={20mm, 15mm, 20mm, 5mm}, width=180mm]{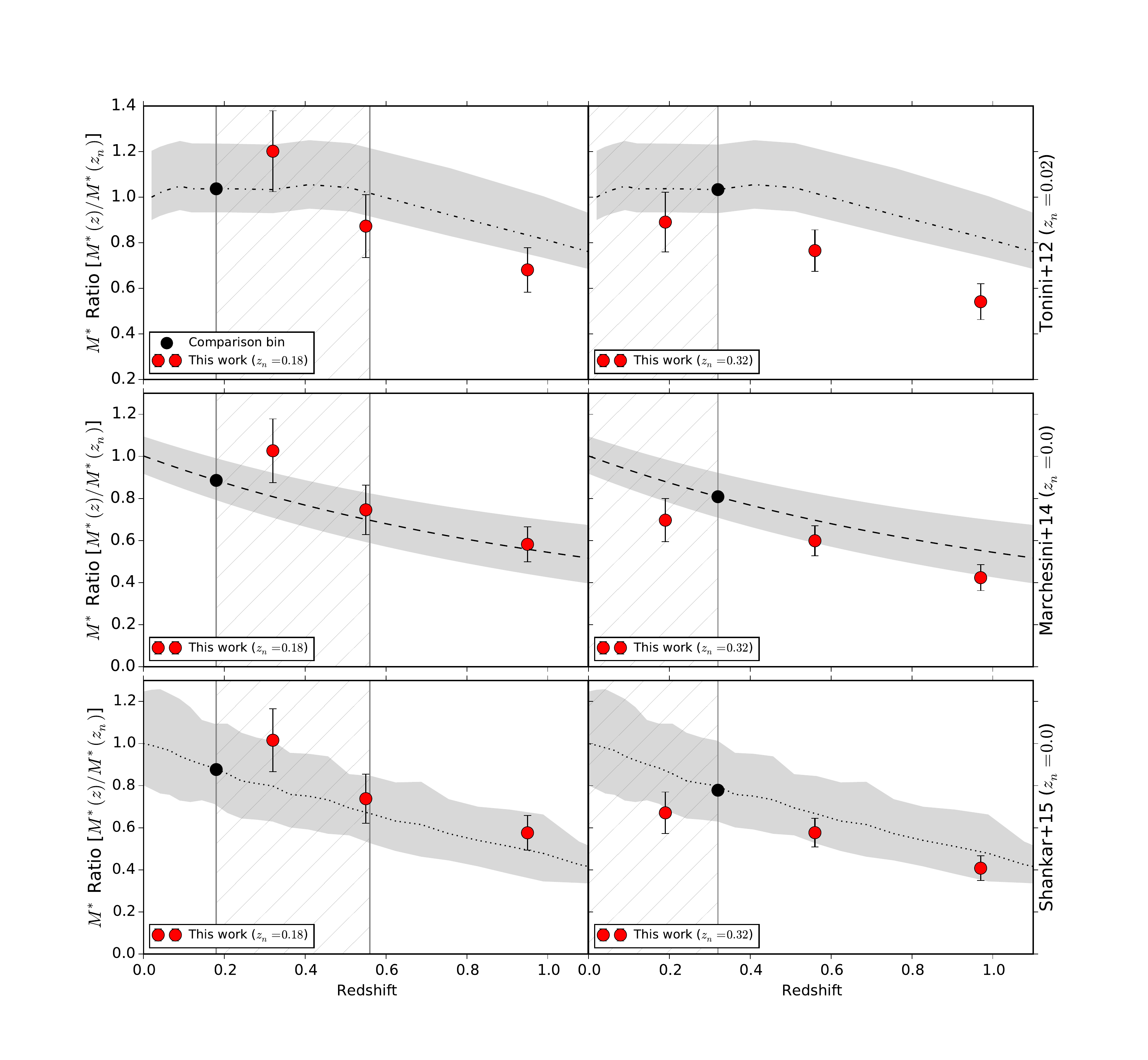}
		\caption{ The evolving stellar mass
                   of BCGs with redshift, presented as the ratio of
                    the $M_{BCG}^*$ at redshift $z$ with the
                    $M_{BCG}^*$ at the comparison redshift $z_n$. The
                    red points indicate the stellar mass ratio of the
                    BCGs in their redshift bin, as compared with the
                    comparison bin (black point). In the left column,
                    the ratios are calculated with a low-redshift
                    comparison bin, whereas the ratios in the right
                    column are calculated with a medium-redshift
                    comparison bin. The results have been scaled such
                    that the comparison bin lies on the model
                   (or on the data, in the case of the data from
                      \citet{Marchesini14}) to which the data are
                    being compared. By changing point at which the
                    difference in stellar mass is calculated, it can
                    be seen that the deviation between data and model
                    may be occurring at either low or medium
                    redshifts, as indicated by the hatched regions.
		     From the top row to the bottom, the
                    three addaitional datasets to which the data are being compared
                    are the
                    simulations model of \citet{Tonini12}; the
                    observational results of \citet{Marchesini14}; and
                    the semi-empirical model of \citet{Shankar15}. }
		\label{fig:MassEvolution}
\end{figure*}

Here we scale our data to the low-redshift,
$0.0<z<0.25$ bin. The results are shown in the left-hand plots of Figures \ref{fig:MassEvolutionSingle} and \ref{fig:MassEvolution}.
In each panel of the graph, the data have been scaled so that the
comparison point lies on the model. 
While the agreement between the data and the DLB07 model is poorer,
the agreement between the data, excluding the intermediate-redshift, $0.25<z<0.40$ bin, and the \citet{Tonini12} model is better.
In this model, the BCG accretes stellar mass earlier than in DLB07.
The data is also in better agreement
with both the observational data from \citet{Marchesini14} and
the semi-empirical model from \citet{Shankar15}.

\subsubsection{The second scaling} 

Here we scale our data to
the intermediate-redshift, $0.25<z<0.40$ bin. The motivation
here is to see if the growth of BCGs stalls at lower redshifts. We first plot this on top of the
predictions made by the DLB07 model, as shown in Figure
\ref{fig:MassEvolutionSingle}.

Although the data match the model very well at higher redshifts, 
it is clear that the low-redshift point lies below the
prediction of the model. We
test the robustness of our results around $z\sim0.2$ by splitting the
low redshift bin into four smaller bins, and compare the masses to the
intermediate redshift bin. While this causes an increase in the errors
of these points due to the smaller sample size of each bin, it is
clear to see from the blue points in the inset of the right panel in
Figure \ref{fig:MassEvolutionSingle} that the mass of
the BCGs in the lower redshift range are consistently lower than what
the DLB07 model predicts. 
This perhaps indicates that the
point at which we tie the observations with the model is
anomalously high, as there is not a smooth transition between this point and the
others at lower redshifts.

To check whether the transition between the high point and those at higher redshifts was smooth, we further split the high redshift bins, as we did for the lower bin. We show these results in the inset within the left panel of Figure \ref{fig:MassEvolutionSingle}. This indicates that in the higher bins, the transition is smooth, indicating that the sharper drop in growth below the intermediate redshift bin may be real.

From the perspective of the data, we
have no strong reasons to suspect that the data in the intermediate-redshift
bin is biased, so we do not choose one interpretation over the
other.  Instead, we discuss the implications of both in the next section.

\begin{table*}
	\centering
	\caption[Mass growth results]{ Results from
          comparing BCGs masses over different redshift
          intervals. Columns (3) and (4) give the median redshifts of
          the lower and upper redshift samples respectively. Column
          (5) then gives the growth factor between the high- and low-z
          samples .The bold font values in either
            column (3) or (4) indicate the redshift bins that were then scaled to the models
            in Figures \ref{fig:MassEvolutionSingle} and Figure \ref{fig:MassEvolution}.}
	\begin{tabular}{ c  c  c  c   c  c  }
	\hline
	\hline
	Low-z Range & High-z Range & Low-z & High-z &  Growth & Clusters per bin\\ 
	(1) & (2) & (3) & (4) & (5) & (6)\\
	\hline
	\multicolumn{5}{l}{\textit{Comparing with the $z=0.18$ bin:}} \\
	0.00-0.25 & 0.25-0.40 & \textbf{0.18} & 0.32 & 0.86$\pm$0.13 & 33\\
	0.00-0.25 & 0.40-0.80 & \textbf{0.18} & 0.55 & 1.19$\pm$0.19 & 26\\
	0.00-0.25 & 0.80-1.60 & \textbf{0.18} & 0.95 & 1.52$\pm$0.22 & 24\\
	\hline
	\multicolumn{5}{l}{\textit{Comparing with the $z=0.32$ bin:}} \\
	0.00-0.25 & 0.25-0.40 & 0.19 & \textbf{0.32} & 0.86$\pm$0.13 & 33\\
	0.25-0.40 & 0.40-0.80 & \textbf{0.32} & 0.56 & 1.35$\pm$0.16 & 27\\
	0.25-0.40 & 0.80-1.60 & \textbf{0.32} & 0.97 & 1.91$\pm$0.28 & 18\\
	\hline
	\multicolumn{5}{l}{\textit{Splitting the low-z bin into four smaller bins:}} \\
	0.00-0.10 & 0.25-0.40 & 0.08 & \textbf{0.32} & 0.88$\pm$0.32 & 7\\
	0.10-0.17 & 0.25-0.40 & 0.15 & \textbf{0.32} & 0.91$\pm$0.20 & 13\\
	0.17-0.21 & 0.25-0.40 & 0.19 & \textbf{0.32} & 0.76$\pm$0.10 & 18\\
	0.21-0.25 & 0.25-0.40 & 0.23 & \textbf{0.32} & 0.91$\pm$0.17 & 14\\
	\hline
	\multicolumn{5}{l}{\textit{Splitting the high-z bins into six smaller bins:}} \\
	0.25-0.40 & 0.40-0.50 & \textbf{0.18} & 0.44 & 1.04$\pm$0.26 & 8\\
	0.25-0.40 & 0.50-0.60 & \textbf{0.18} & 0.54 & 1.09$\pm$0.28 & 9\\
	0.25-0.40 & 0.60-0.70 & \textbf{0.18} & 0.65 & 1.91$\pm$0.48 & 7\\
	0.25-0.40 & 0.70-0.80 & \textbf{0.18} & 0.73 & 1.08$\pm$0.44 & 4\\
	0.25-0.40 & 0.80-1.00 & \textbf{0.18} & 0.88 & 1.50$\pm$0.36 & 14\\
	0.25-0.40 & 1.00-1.20 & \textbf{0.18} & 1.1 & 1.38$\pm$0.30 & 6\\
	\hline
	\end{tabular}
	\label{Table:GrowthResults}
\end{table*}

\section{Discussion}
\label{sec:Discussion}

\subsection{Systematic Uncertainties in the Analysis}

The points in Figures
\ref{fig:MassEvolutionSingle} and \ref{fig:MassEvolution}, were computed using the BC03 stellar population synthesis model and a Chabrier initial mass function (IMF). However, some of the models, plotted as dotted lines in Figure \ref{fig:MassEvolution}, use different stellar population synthesis models. For example, the models of \citet{Tonini12} use the models from \citet{Maraston05}, which have a strong post-AGB phase. Hence, we need to make sure that the differences between the models and the data are not driven by some of these differences.

We compared a number of models, e.g. models that use a Chabrier IMF with models that use a Salpeter IMF, and models from  \citet{Maraston05} with BC03 models.
We find that, while the ratio in the stellar masses of the two models can vary by 25\% from
the redshift of formation to today, this ratio varies by only a few percent over the redshift range of interest in this paper (i.e. $z\sim1.6$ to $z\sim0$). Hence, the differences in Figure \ref{fig:MassEvolution} between the data and the models are not driven by the assumptions that went into the models.

We additionally checked to see whether the selection method of the clusters has an impact on the measured growth factor. To do this, we measured the growth over two bins ($0<z<0.4$ and $0.4<z<1.6$) for all the data, and then individually for those BCGs from X-ray selected clusters, and also those from Sunyaev-Zel'Dovich selected clusters. We note that the results all agree within error, and that there is therefore no bias as a result of selection. Due to low numbers of optically selected clusters at high redshift, we were unable to carry out this same check for this method of selection. 

\subsection{The Growth of Clusters and their BCG}

\begin{figure}
	\includegraphics[trim=8mm 5mm 8mm 0mm, width=84mm]{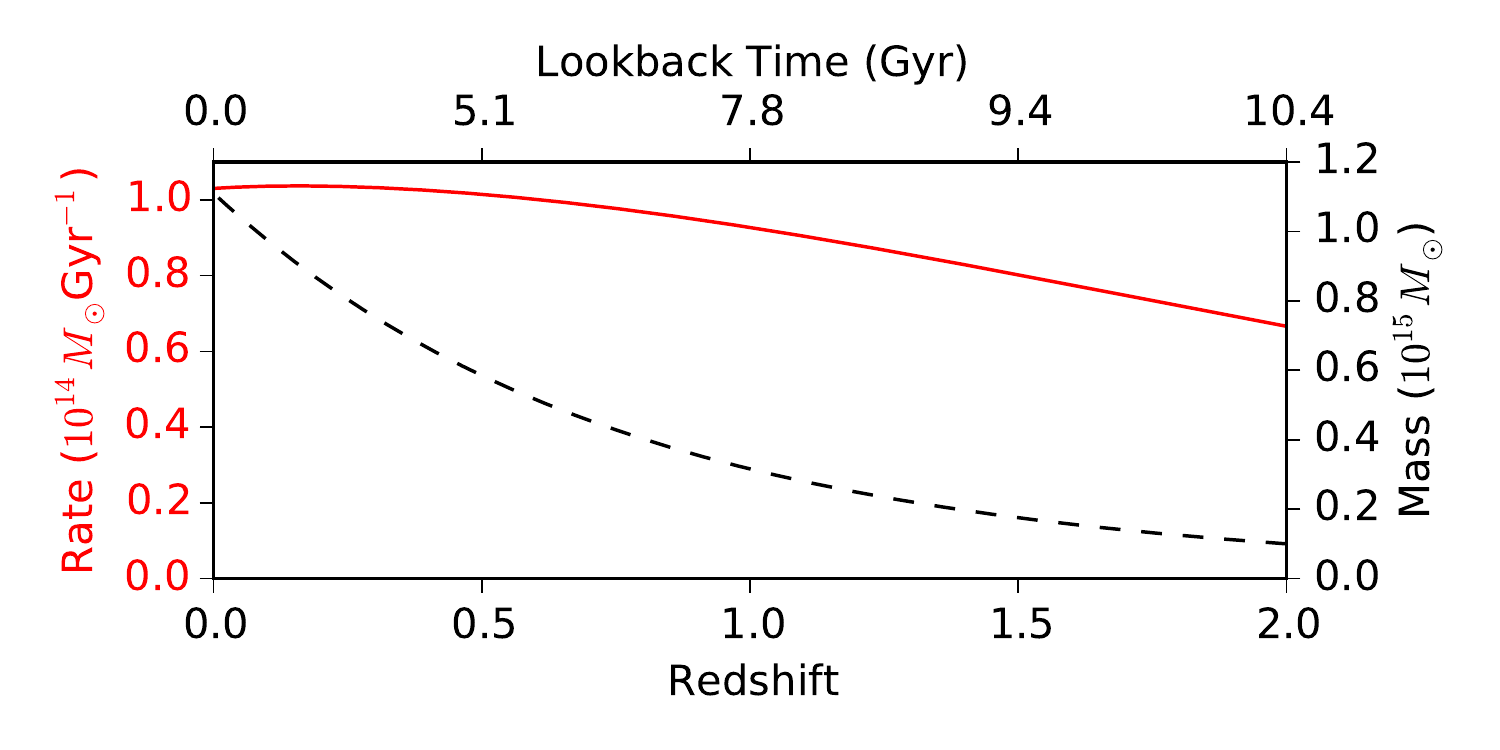}
		\caption{The mass (dashed) and mass accretion rate
                  (red) of a galaxy cluster with a mass of $1\times
                  10^{14} M_{\odot}$ at $z=2$, according to the mean
                  cluster accretion rate equation of
                  \citet{Fakhouri10}. Note that the cluster accretion
                  rate is roughly constant over time, especially since
                  $z=0.5$. }
		\label{fig:ClusterMassAccretion}
\end{figure}

In the flat $\Lambda$CDM model, a $10^{15} M_{\odot}$ cluster at $z=0$ is
expected to have grown by a factor of about 10 since $z=2$, as shown in Figure
\ref{fig:ClusterMassAccretion}. Since $z=1$, it is expected that the cluster has grown
by a factor of about 3. Observationally, we find that the stellar masses
of BCGs grow by a factor of about 2. Hence, over this redshift
interval, the cluster grows more quickly than the central BCG. 
There are a couple of reasons for this.

Firstly, as the cluster grows and becomes larger, the timescale for infalling
galaxies to reach the BCG increases. Galaxies that enter the cluster
via the outskirts will feel a frictional drag (termed dynamical
friction) from the wake they create as they move in the cluster. Over
time, dynamical friction brings them closer to the core, and at some
point they will merge with the central galaxy. The timescale of this
process is \citep{Binney87}:
\begin{equation}
t_{\mbox{fric}}=\frac{2.34}{\ln\Lambda}\frac{\sigma_M^2}{\sigma_S^3}r_i
\label{eqn:DynamicalFriction}
\end{equation}
where $\sigma_M$ is the velocity dispersion of the central galaxy, and $\sigma_S$ is the velocity dispersion of the satellite; 
$r_i$ represents the radius from which the satellite is spiralling in, and ln $\Lambda$ is the Coulomb logarithm. 

For a satellite of a given mass, the timescale is proportional to the
halo mass to the two-thirds power and to the radius of the cluster,
both of which are increasing in time. The net effect is that it takes
longer for material entering the cluster to reach its centre. This,
combined with the relatively flat accretion rate (see Figure
\ref{fig:ClusterMassAccretion}), results in less material reaching the
cluster centre as the cluster gets bigger.

The dynamical friction timescale is explicitly included in models that study
the growth of BCGs \citep[DLB07,][]{Shankar15}. Differences in the growth rate of BCGs
between models and the data may be due to the way the dynamical friction formula
is used in the models. By reducing this time scale by one-third, \citet{Shankar15} was able
to increase the amount of stellar material accreted by BCGs from $z=1$ to today by $\sim20$\%.

Secondly, as an infalling satellite moves through the
cluster, tidal stripping as a result of interactions with
surrounding objects will occur. As clusters get larger, the number
of interactions an infalling satellite will experience will also
increase. If the amount of stellar stripping experienced by the
infalling satellite is the same per interaction, then by the time
the satellite has fallen sufficiently far into the cluster to merge
with the BCG, the amount of remaining stellar mass within the
satellite available to be accreted by the BCG is less in more massive clusters.
Some models set the amount of stripping to zero
\citep{Shankar15}; others do not include it all. Hence, discrepancies between the
data and the models may be due to the way tidal striping is included in the model.

Other effects, not captured fully in the data, may contribute
to differences between models and the data. In recent work,
\citet{Burke15} have shown that the intra-cluster light (ICL) grows
substantially below a redshift of $z\sim0.4$.  It therefore seems
reasonable to posit that most of the mass that reaches the cluster
core below $z=0.3$ ends up in the ICL and not the BCG. As local BCGs
are already quite large, most mergers that occur would be minor
mergers. Since minor mergers do not affect the inner cores of the
progenitor galaxies, the infalling stars would be inclined to stay on
the outskirts of the BCG, and hence contribute to the ICL.

Measuring the amount of material in the ICL is a challenging observation,
especially in the K band, where it is several orders of magnitude
fainter than the night sky. 
As mentioned in Section \ref{sec:photometry}, {\tt MAG\_AUTO} does not recover the full amount of
galaxy light, indicating that light in the outskirts of the galaxy is being neglected by the aperture treatment of {\tt MAG\_AUTO}. 
An alternative approach to reconciling
observations with simulations may be to extend the simulations, so
that one creates simulated images, as has been recently done in the
Illustris simulation\footnote{http://www.illustris-project.org}. This
would enable one to make the exact same measurement on the simulated
image and the real data, thus circumventing some of the biases in the
comparison.

In the previous section, we found that we had good
  agreement with more recent models in the literature \citep[for
    example,][]{Shankar15}, if we anchored the data to the
  models at the low-redshift end. One would then interpret the excess
  in the intermediate-redshift bin as a slight anomaly.

If one instead anchors the data to the models at the
  intermediate-redshift bin, we find that we have poorer agreement
  with the more recent models and better agreement with the DLB07
  model, but only in the higher redshift bins. The low redshift point
  (see right panel of Figure \ref{fig:MassEvolutionSingle}) falls well short of the model.

We do not discount either interpretation, as there are no
  reasons to believe that the clusters in the intermediate-redshift
  bin lead to a biased measurement in that bin. However, if the latter
  interpretation is correct, it would mean that the stellar material that
  contributes to the build-up of mass predicted by the DLB07 model at
  late times must be done in such a way that it and some of the
  stellar material that is already in the BCG are distributed outside
  the apertures that we use to measure fluxes in the Ks-band. This can
  only happen if the profile of the BCG changes as well.

\section{Summary and Conclusions}
\label{sec:Conclusion}

We have added a sample of 102 BCGs with known cluster masses 
to an existing sample of 155 BCGs to create a 
BCG sample spanning
the redshift range $0.04 < z < 1.63$. We use this sample to study the
stellar mass growth of BCGs.

We find that the build-up of stellar mass of BCGs from $z\sim1$ to today, as inferred from the observer-frame
Ks band, is broadly consistent with predictions from recent semi-analytic and semi-empirical models. 

The BCGs in the very lowest redshift bin have a lower stellar mass 
than the median-redshift bin, providing tentative evidence that the stellar mass growth rate of BCGs may be slowing. 

In order to better constrain the growth rate at lower redshifts, it will be necessary to increase the number of BCGs and
to better match the methods used to to derive masses from observations and theoretical models.

\section{Acknowledgements}

The authors would like to thank the referee for his
constructive comments, and whose suggestions have helped make this study a more
cohesive analysis. 
SB acknowledges the help and suggestions of S. Brough and
P. Oliva-Altamirano, and the support of the AAO Honours/Masters Scholarship. 
DM, ZCM, CW, and NK acknowledge the support of
the Research Corporation for Science Advancement’s Cottrell
Scholarship. 
JvdS is funded under Bland-Hawthorn’s ARC Laureate Fellowship (FL140100278).
ZCM gratefully acknowledges support from the John
F. Burlingame and the Kathryn McCarthy Graduate Fellowships in Physics
at Tufts University.

This work is based in part on observations taken at the European
Southern Observatory New Technology Telescope (ESO programs 092.A-0857
and 094.A-0531), and also on observations taken with the WYIN
telescope at the Kitt Peak National Observatory. Additionally, we
made use of the ESO Science Archive Facility.

We have used Python, in particular
the packages numpy, scipy and astropy, for the data analysis, and
matplotlib \citep{Hunter07} for the generation of the plots used
within this paper.

%\appendix

\begin{landscape}
\begin{table}
	\caption[BCG Results]{BCG results. \\
Redshift sources as given in Table \ref{ObservationalSummary} \\ 
Cluster mass proxy sources: (1) \citet{Mann12} (2) \citet{Mantz10} (3) \citet{Menanteau13} (4) \citet{Ebeling07} (5) \citet{Menanteau10} (6) \citet{Mahdavi13} (7) \citet{Piffaretti11} (8) \citet{Henry92} (9) \citet{Ebeling98} (10) \citet{Maughan12} (11) \citet{Benson13} (12) \citet{Buddendiek15} (13) \citet{Hasselfield13}\\ 
* Magnitude errors stated are those reported by {\sc SExtractor}. \\
$\star$ BCG masses are uncertain by $\sim20\%$.\\
$\dagger$ All cluster masses are measured to $M_{200}$.}
	\label{BCGClusterResults}
	\centering
	\begin{tabular}{@{}lcccclccccc}
		\hline
		\hline
		Cluster & $z_{spec}$ & $z_{phot}$ & z-Source & RA & Dec & BCG Ks-Mag * & BCG Mass $\star$  & Cluster Mass $\dagger$ & Cluster Mass Proxy & Source\\
		  &  &  &  & & & & $\times 10^{12} M_{\odot}$  &  $\times 10^{15} M_{\odot}$ & & \\
		\hline
SPT-CL-J0000-5748 & 0.702 & ... & (20) & 00:01:0.060 & -57:48:33.43 & $15.047 \pm 0.017$ & $1.065$ & $0.59 \pm 0.10$ & SZ Y Parameter & (11)\\
MACS-J0011.7-1523 & 0.379 & ... & (18) & 00:11:42.82 & -15:23:21.07 & $14.914 \pm 0.004$ & $0.355$ & $1.00 \pm 0.25$ & X-ray Luminosity & (1)\\
MACS-J0014.3-3022 & 0.308 & ... & (1) & 00:14:15.82 & -30:22:14.4 & $12.669 \pm 0.003$ & $1.743$ & $2.57 \pm 0.34$ & Gas Mass & (2)\\
MACS-J0014.3-3022 & 0.308 & ... & (1) & 00:14:17.26 & -30:22:34.8 & $12.927 \pm 0.002$ & $1.375$ & $2.57 \pm 0.34$ & Gas Mass & (2)\\
ACT-CL-J0014.9-0057 & 0.533 & ... & (7) & 00:14:54.10 & -00:57:7.470 & $15.081 \pm 0.013$ & $0.627$ & $1.28 \pm 0.74$ & X-ray Luminosity & (3)\\
C1G-J001640.6-130644 & ... & 0.700 & (21) & 00:16:40.71 & -13:06:43.7 & $16.160 \pm 0.220$ & $0.484$ & $1.55 \pm 1.06$ & X-ray Luminosity & (12)\\
ACT-CL-J0017.6-0051 & 0.211 & ... & (7) & 00:17:37.61 & -00:52:41.79 & $13.796 \pm 0.008$ & $0.301$ & $0.21 \pm 0.15$ & X-ray Luminosity & (3)\\
ACT-CL-J0018.2-0022 & ... & 0.75 & (3) & 00:18:14.21 & -00:22:32.18 & $16.066 \pm 0.029$ & $0.469$ & $0.45 \pm 0.13$ & S-Z Parameter & (13)\\
MACS-J0025.4-1222 & 0.5843 & ... & (1) & 00:25:27.42 & -12:22:22.76 & $15.896 \pm 0.017$ & $0.351$ & $0.92 \pm 0.50$ & X-ray Temperature & (4)\\
MACS-J0025.4-1222 & 0.5843 & ... & (1) & 00:25:33.01 & -12:23:16.26 & $15.50 \pm 0.014$ & $0.505$ & $0.92 \pm 0.50$ & X-ray Temperature & (4)\\
MACS-J0032.1+1808 & ... & 0.398 & (22) & 00:32:9.380 & +18:06:56.17 & $14.560 \pm 0.013$ & $0.569$ & ... & ... & ...\\
MACS-J0033.8-0751 & ... & 0.3 & (22) & 00:33:53.14 & -07:52:10.13 & $14.470 \pm 0.011$ & $0.332$ & ... & ... & ...\\
MACS-J0034.4+0225 & ... & 0.35 & (22) & 00:34:28.15 & +02:25:22.84 & $15.020 \pm 0.004$ & $0.275$ & ... & ... & ...\\
MACS-J0034.9+0234 & ... & ... & ... & 00:34:58.03 & +02:33:33.53 & $14.159 \pm 0.004$ & ... & ... & ... & ...\\
MACS-J0035.4-2015 & 0.352 & ... & (1) & 00:35:26.20 & -20:15:42.19 & $14.165 \pm 0.005$ & $0.606$ & $0.93 \pm 0.23$ & X-ray Luminosity & (1)\\
Abell68 & 0.255 & ... & (12) & 00:37:06.87 & +09:09:25.50 & $13.418 \pm 0.004$ & $0.625$ & $1.11 \pm 0.17$ & Gas Mass & (2)\\
SMACS-J0040.8-4407 & ... & 0.4 & (14) & 00:40:49.97 & -44:07:50.2 & $14.414 \pm 0.006$ & $0.651$ & ... & ... & ...\\
C1G-J005805.6+003058 & 0.662 & ... & (21) & 00:58:05.71 & +00:30:58.2 & $16.565 \pm 0.220$ & $0.232$ & $1.41 \pm 0.92$ & X-ray Luminosity & (12)\\
ACT-CL-J0104.8+0002 & 0.277 & ... & (7) & 01:04:55.35 & +00:03:36.41 & $13.749 \pm 0.004$ & $0.547$ & $0.51 \pm 0.26$ & X-ray Luminosity & (3)\\
MACS-J0110.1+3211 & ... & ... & ... & 01:10:7.190 & +32:10:48.68 & $14.335 \pm 0.004$ & ... & ... & ... & ...\\
ACT-CL-J0119.9+0055 & ... & 0.72 & (3) & 01:19:58.15 & +00:55:34.03 & $16.075 \pm 0.004$ & $0.44$ & $0.48 \pm 0.12$ & S-Z Parameter & (13)\\
ACT-CL-J0127.2+0020 & 0.379 & ... & (7) & 01:27:16.64 & +00:20:41.18 & $14.363 \pm 0.005$ & $0.59$ & $0.77 \pm 0.41$ & X-ray Luminosity & (3)\\
C1G-J013710.4-103423 & ... & 0.575 & (21) & 01:37:10.55 & -10:34:22.1 & $15.265 \pm 0.220$ & $0.625$ & $0.93 \pm 0.64$ & X-ray Luminosity & (12)\\
RX-J0142.0-2131 & ... & 0.28 & (7) & 01:42:03.42 & +21:31:17.0 & $14.280 \pm 0.007$ & $0.336$ & ... & ... & ...\\
ACT-CL-J0145-5301 & 0.118 & ... & (15) & 01:45:3.590 & -53:01:23.13 & $12.798 \pm 0.016$ & $0.332$ & $0.57 \pm 0.21$ & X-ray Luminosity & (5)\\
MACS-J0150.3-1005 & 0.365 & ... & (12) & 01:50:21.27 & -10:05:30.19 & $13.730 \pm 0.006$ & $1.057$ & ... & ... & ...\\
ACT-CL-J0152.7+0100 & 0.23 & ... & (7) & 01:52:41.97 & +01:00:26.01 & $12.964 \pm 0.012$ & $0.791$ & $0.75 \pm 0.30$ & X-ray Luminosity & (3)\\
ACT-CL-J0156.4-0123 & ... & 0.45 & (3) & 01:56:24.29 & -01:23:17.28 & $14.055 \pm 0.003$ & $1.176$ & $0.45 \pm 0.13$ & S-Z Parameter & (13)\\
ACT-CL-J0206.2-0114 & 0.676 & ... & (7) & 02:06:13.14 & -01:14:59.94 & $16.432 \pm 0.004$ & $0.28$ & $2.13 \pm 1.16$ & X-ray Luminosity & (3)\\
ACT-CL-J0215-5212 & 0.48 & ... & (17) & 02:15:12.33 & -52:12:25.4 & $15.177 \pm 0.012$ & $0.467$ & $0.90 \pm 0.69$ & X-ray Luminosity & (5)\\
ACT-CL-J0223.1-0056 & 0.663 & ... & (7) & 02:15:28.49 & +00:30:37.52 & $15.971 \pm 0.003$ & $0.428$ & $0.41 \pm 0.12$ & S-Z Parameter & (13)\\
ACT-CL-J0218.2-0041 & 0.672 & ... & (3) & 02:18:17.61 & -00:41:38.73 & $16.320 \pm 0.006$ & $0.31$ & $0.55 \pm 0.12$ & S-Z Parameter & (13)\\
ACT-CL-J0219.8+0022 & 0.537 & ... & (3) & 02:19:50.42 & +00:22:14.7 & $15.689 \pm 0.220$ & $0.358$ & $0.44 \pm 0.13$ & S-Z Parameter & (13)\\
ACT-CL-J0219.9+0129 & ... & 0.35 & (3) & 02:19:52.16 & +01:29:52.4 & $14.432 \pm 0.220$ & $0.473$ & $0.66 \pm 0.46$ & X-ray Luminosity & (3)\\
ACT-CL-J0221.5-0012 & 0.589 & ... & (7) & 02:21:36.51 & -00:12:22.37 & $15.902 \pm 0.025$ & $0.375$ & $1.08 \pm 0.86$ & X-ray Luminosity & (3)\\
ACT-CL-J0232-5257 & 0.556 & ... & (17) & 02:32:42.72 & -52:57:22.59 & $15.072 \pm 0.022$ & $0.691$ & $0.98 \pm 0.71$ & X-ray Luminosity & (5)\\
ACT-CL-J0232-5257 & 0.556 & ... & (17) & 02:32:49.42 & -52:57:11.48 & $15.349 \pm 0.019$ & $0.536$ & $0.98 \pm 0.71$ & X-ray Luminosity & (5)\\
ACT-CL-J0235-5121 & 0.278 & ... & (17) & 02:35:45.26 & -51:21:4.770 & $13.143 \pm 0.035$ & $0.956$ & $1.20 \pm 0.56$ & X-ray Luminosity & (5)\\
ACT-CL-J0237-4939 & 0.334 & ... & (17) & 02:37:1.670 & -49:38:9.680 & $14.063 \pm 0.014$ & $0.567$ & $0.59 \pm 0.33$ & X-ray Luminosity & (5)\\
Abell370 & 0.375 & ... & (7,12) & 02:39:51.90 & -01:35:14.72 & $14.254 \pm 0.009$ & $0.652$ & $0.89 \pm 0.68$ & X-ray Luminosity & (3)\\
Abell370 & 0.375 & ... & (7,12) & 02:39:52.31 & -01:35:52.15 & $13.987 \pm 0.008$ & $0.834$ & $0.89 \pm 0.68$ & X-ray Luminosity & (3)\\
MACS-J0242.5-2132 & 0.314 & ... & (1) & 02:42:35.91 & -21:32:25.8 & $13.614 \pm 0.005$ & $0.857$ & $1.12 \pm 0.19$ & Gas Mass & (2)\\
		\hline
	\end{tabular}
\end{table}
\end{landscape}

\begin{landscape}
\begin{table}
	\centering
	\begin{tabular}{@{}lcccclccccc}
		\hline
		\hline
		Cluster & $z_{spec}$ & $z_{phot}$ & z-Source & RA & Dec & BCG Ks-Mag * & BCG Mass $\star$  & Cluster Mass & Cluster Mass Proxy & Source\\
		  &  &  &  & & & & $\times 10^{12} M_{\odot}$  &  $\times 10^{15} M_{\odot}$ & & \\
		\hline
ACT-CL-J0245.8-0042 & 0.179 & ... & (8) & 02:45:51.74 & -00:42:16.34 & $13.372 \pm 0.016$ & $0.354$ & $0.25 \pm 0.21$ & X-ray Luminosity & (3)\\
ACT-CL-J0250.1+0008 & ... & 0.78 & (3) & 02:50:8.40 & +00:08:16.22 & $16.360 \pm 0.003$ & $0.38$ & $0.39 \pm 0.12$ & S-Z Parameter & (13)\\
ACT-CL-J0256+0006 & 0.363 & ... & (8) & 02:56:30.84 & +00:06:03.3 & $14.747 \pm 0.008$ & $0.414$ & $1.06 \pm 0.45$ & X-ray Luminosity & (3)\\
ACT-CL-J0256+0006 & 0.363 & ... & (8) & 02:56:33.76 & +00:06:28.8 & $15.067 \pm 0.011$ & $0.308$ & $1.06 \pm 0.45$ & X-ray Luminosity & (3)\\
MACS-J0257.1-2325 & 0.505 & ... & (1) & 02:57:8.760 & -23:26:4.890 & $14.686 \pm 0.007$ & $0.817$ & $2.18 \pm 1.19$ & X-ray Temperature & (4)\\
MACS-J0257.1-2325 & 0.505 & ... & (1) & 02:57:8.790 & -23:26:4.930 & $14.510 \pm 0.002$ & $0.96$ & $2.18 \pm 1.19$ & X-ray Temperature & (4)\\
ACT-CL-J0301.6+0155 & 0.167 & ... & (9) & 03:01:38.20 & +01:55:14.66 & $13.084 \pm 0.006$ & $0.461$ & $0.54 \pm 0.25$ & X-ray Luminosity & (3)\\
ACT-CL-J0304-4921 & 0.392 & ... & (17) & 03:04:16.18 & -49:21:26.3 & $14.263 \pm 0.007$ & $0.748$ & $1.22 \pm 0.58$ & X-ray Luminosity & (5)\\
SMACS-J0304.3-4401 & ... & 0.52 & (17) & 03:04:16.86 & -44:01:31.5 & $15.035 \pm 0.010$ & $0.654$ & ... & ... & ...\\
SMACS-J0304.3-4401 & ... & 0.52 & (17) & 03:04:21.09 & -44:02:37.51 & $15.064 \pm 0.014$ & $0.637$ & ... & ... & ...\\
ACT-CL-J0326.8-0043 & 0.448 & ... & (9) & 03:26:49.94 & -00:43:51.61 & $14.339 \pm 0.007$ & $0.905$ & $1.46 \pm 0.75$ & X-ray Luminosity & (3)\\
ACT-CL-J0330-5227 & 0.44 & ... & (9,10) & 03:30:56.96 & -52:28:13.2 & $14.274 \pm 0.008$ & $0.958$ & $2.86 \pm 0.97$ & X-ray Luminosity & (5)\\
ACT-CL-J0346-5438 & 0.53 & ... & (17) & 03:46:55.48 & -54:38:55.0 & $15.063 \pm 0.014$ & $0.637$ & $0.99 \pm 0.54$ & X-ray Luminosity & (5)\\
ACT-CL-J0348+0029 & 0.297 & ... & (17) & 03:48:36.71 & +00:29:32.9 & $14.496 \pm 0.009$ & $0.324$ & $0.50 \pm 0.30$ & X-ray Luminosity & (3)\\
ACT-CL-J0348-0028 & 0.345 & ... & (17) & 03:48:38.25 & -00:28:08.6 & $13.869 \pm 0.003$ & $0.795$ & $0.69 \pm 0.40$ & X-ray Luminosity & (3)\\
MACS-J0358.8-2955 & 0.425 & ... & (1) & 03:58:54.09 & -29:55:30.8 & $14.415 \pm 0.016$ & $0.746$ & $2.3 \pm 0.67$ & Gas Mass & (2)\\
MACS-J0416.1-2403 & 0.40 & ... & (2) & 04:16:09.15 & -24:04:02.1 & $14.084 \pm 0.007$ & $0.882$ & $0.97 \pm 0.24$ & X-ray Luminosity & (1)\\
MACS-J0417.5-1154 & 0.443 & ... & (2) & 04:17:34.49 & -11:54:34.3 & $13.877 \pm 0.004$ & $1.385$ & $3.22 \pm 0.39$ & Gas Mass & (2)\\
Abell496 & 0.0328 & ... & (7) & 04:33:37.84 & -13:15:43.10 & $10.743 \pm 0.001$ & $0.185$ & $0.43 \pm 0.11$ & X-ray Luminosity & (7)\\
ACT-CL-J0438-5419 & 0.421 & ... & (17) & 04:38:17.68 & -54:19:20.5 & $14.045 \pm 0.005$ & $1.049$ & $2.75 \pm 1.16$ & X-ray Luminosity & (5)\\
SMACS-J0439-4600 & ... & ... & ... & 04:39:13.91 & -46:00:48.55 & $14.271 \pm 0.010$ & ... & ... & ... & ...\\
MS-0440+0204 & 0.197 & ... & (5) & 04:43:16.14 & +02:10:02.4 & $13.180 \pm 0.002$ & $0.53$ & $0.44 \pm 0.11$ & X-ray Luminosity & (8)\\
MS-0440+0204 & 0.197 & ... & (5) & 04:43:16.29 & +02:10:04.6 & $13.230 \pm 0.001$ & $0.506$ & $0.44 \pm 0.11$ & X-ray Luminosity & (8)\\
MACS-J0449-2848 & ... & ... & ... & 04:49:20.76 & -28:49:08.19 & $14.916 \pm 0.009$ & ... & ... & ... & ...\\
MACS-J0451.6-0305 & 0.5386 & ... & (5) & 04:54:10.84 & -03:00:51.5 & $15.691 \pm 0.015$ & $0.39$ & $1.22 \pm 0.30$ & X-ray Luminosity & (1)\\
MACS-J0454.1-0300 & 0.5377 & ... & (6) & 04:54:16.12 & -02:59:26.4 & $15.367 \pm 0.008$ & $0.525$ & $1.06 \pm 0.58$ & X-ray Temperature & (4)\\
Abell520 & 0.205 & ... & (2) & 04:54:3.810 & +02:53:32.05 & $13.497 \pm 0.028$ & $0.396$ & $1.04 \pm 0.26$ & X-ray Luminosity & (1)\\
ACT-CL-J0516-5430 & 0.294 & ... & (23) & 05:16:37.35 & -54:30:1.520 & $13.30 \pm 0.002$ & $0.974$ & $1.47 \pm 0.57$ & X-ray Luminosity & (5)\\
SPT-CL-J0553-5005 & 0.881 & ... & (24) & 05:33:37.50 & -50:06:4.750 & $16.826 \pm 0.010$ & $0.305$ & $0.37 \pm 0.08$ & SZ Y Parameter & (11)\\
SPT-CL-J0546-5345 & 1.066 & ... & (27) & 05:46:37.66 & -53:45:31.08 & $17.062 \pm 0.002$ & $0.333$ & $2.13 \pm 1.37$ & X-ray Luminosity & (5)\\
MACS-J0553.4-3342 & 0.407 & ... & (12) & 05:53:19.35 & -33:42:27.4 & $15.228 \pm 0.005$ & $0.308$ & ... & ... & ...\\
MACS-J0553.4-3342 & 0.407 & ... & (12) & 05:53:25.77 & -33:42:28.0 & $14.534 \pm 0.010$ & $0.583$ & ... & ... & ...\\
SMACS-J0600.2-2008 & ... & 0.46 & (12) & 06:00:08.18 & -20:08:09.0 & $14.965 \pm 0.008$ & $0.509$ & ... & ... & ...\\
SMACS-J0600-4353 & ... & ... & ... & 06:00:13.05 & -43:53:30.47 & $14.305 \pm 0.003$ & ... & ... & ... & ...\\
SMACS-J0600.2-2008 & ... & 0.46 & (12) & 06:00:16.83 & -20:06:55.7 & $14.614 \pm 0.008$ & $0.702$ & ... & ... & ...\\
ACT-CL-J0616-5227 & 0.684 & ... & (17) & 06:16:33.92 & -52:27:09.9 & $15.001 \pm 0.011$ & $1.045$ & $1.88 \pm 0.92$ & X-ray Luminosity & (5)\\
ACT-CL-J0641-4949 & 0.146 & ... & (9,11) & 06:41:37.81 & -49:46:54.5 & $12.735 \pm 0.003$ & $0.487$ & $0.18 \pm 0.10$ & X-ray Luminosity & (5)\\
ACT-CL-J0645-5413 & 0.167 & ... & (15) & 06:45:29.48 & -54:13:36.93 & $12.624 \pm 0.005$ & $0.704$ & $1.13 \pm 0.36$ & X-ray Luminosity & (5)\\
ACT-CL-J0707-5522 & 0.296 & ... & (26) & 07:07:04.70 & -55:23:08.5 & $13.368 \pm 0.005$ & $0.916$ & $1.24 \pm 0.55$ & X-ray Luminosity & (5)\\
SMACS-J0723.3-7327 & ... & 0.39 & (16) & 07:23:18.46 & -73:27:17.0 & $15.062 \pm 0.043$ & $0.358$ & ... & ... & ...\\
C1G-J080434.9+330509 & 0.552 & ... & (21) & 08:04:35.12 & +33:05:8.430 & $15.032 \pm 0.004$ & $0.717$ & $0.91 \pm 0.59$ & X-ray Luminosity & (12)\\
Abell644 & 0.0704 & ... & (7) & 08:17:25.62 & -07:30:45.40 & $11.818 \pm 0.002$ & $0.27$ & $0.57 \pm 0.14$ & X-ray Luminosity & (7)\\
MACSJ-0845.4+0327 & ... & ... & ... & 08:45:27.76 & +03:27:38.93 & $14.106 \pm 0.030$ & ... & ... & ... & ...\\
MACS-J0850.1+3604 & 0.378 & ... & (19) & 08:50:7.840 & +36:04:11.48 & $14.356 \pm 0.037$ & $0.594$ & ... & ... & ...\\
Abell750 & 0.18 & ... & (12) & 09:09:12.77 & +10:58:28.72 & $13.274 \pm 0.003$ & $0.388$ & ... & ... & ...\\
MACS-J0911.2+1746 & 0.5049 & ... & (6) & 09:11:11.55 & +17:46:28.7 & $15.111 \pm 0.016$ & $0.552$ & $1.51 \pm 0.82$ & X-ray Temperature & (4)\\
C1G-J094811.6+290709 & 0.778 & ... & (21) & 09:48:11.50 & +29:07:12.15 & $16.158 \pm 0.007$ & $0.458$ & $2.02 \pm 1.18$ & X-ray Luminosity & (12)\\
MACS-J0949.8+1708 & 0.384 & ... & (1) & 09:49:51.81 & +17:07:9.580 & $14.326 \pm 0.008$ & $0.612$ & $1.65 \pm 0.34$ & Gas Mass & (2)\\
Abell901 & 0.170 & ... & (1) & 09:56:29.92 & -10:05:42.26 & $14.521 \pm 0.007$ & $0.123$ & ... & ... & ...\\
Abell963 & 0.206 & ... & (1) & 10:17:3.660 & +39:02:49.71 & $12.651 \pm 0.006$ & $0.863$ & $0.87 \pm 0.47$ & X-ray Temperature & (6)\\
MACS-J1105.7-1014 & 0.466 & ... & (1) & 11:05:46.82 & -10:14:46.1 & $14.491 \pm 0.008$ & $0.878$ & $0.83 \pm 0.20$ & X-ray Luminosity & (1)\\
		\hline
	\end{tabular}
\end{table}
\end{landscape}

\begin{landscape}
\begin{table}
	\centering
	\begin{tabular}{@{}lcccclccccc}
		\hline
		\hline
		Cluster & $z_{spec}$ & $z_{phot}$ & z-Source & RA & Dec & BCG Ks-Mag * & BCG Mass $\star$  & Cluster Mass & Cluster Mass Proxy & Source\\
		  &  &  &  & & & & $\times 10^{12} M_{\odot}$  &  $\times 10^{15} M_{\odot}$ & & \\
		\hline
MACS-J1108.8+0906 & 0.466 & ... & (19) & 11:08:55.34 & +09:06:3.160 & $14.685 \pm 0.001$ & $0.734$ & $0.80 \pm 0.20$ & X-ray Luminosity & (1)\\
Abell1204 & 0.171 & ... & (12) & 11:13:20.51 & +17:35:41.00 & $13.40 \pm 0.004$ & $0.344$ & $0.92 \pm 0.23$ & X-ray Luminosity & (9)\\
MACS-J1115.8+0129 & 0.355 & ... & (12) & 11:15:51.93 & +01:29:55.2 & $14.405 \pm 0.009$ & $0.487$ & $1.26 \pm 0.17$ & Gas Mass & (2)\\
MACS-J1206.2-0847 & 0.439 & ... & (1) & 12:06:12:16 & -08:48:03.1 & $14.618 \pm 0.007$ & $0.698$ & $2.80 \pm 0.44$ & Gas Mass & (2)\\
Abell1553 & 0.165 & ... & (7) & 12:30:48.86 & +10:32:47.30 & $12.385 \pm 0.003$ & $0.877$ & $0.92 \pm 0.23$ & X-ray Luminosity & (9)\\
Abell1634 & 0.196 & ... & (12) & 12:54:01.85 & -06:42:14.00 & $13.481 \pm 0.006$ & $0.401$ & $0.6 0\pm 0.15$ & X-ray Luminosity & (7)\\
MACS-J1311.0-0311 & 0.494 & ... & (12) & 13:11:01.77 & -03:10:40.5 & $14.977 \pm 0.010$ & $0.625$ & $0.98 \pm 0.24$ & X-ray Luminosity & (1)\\
Abell1689 & 0.183 & ... & (7) & 13:11:27.20 & -01:18:45.50 & $13.136 \pm 0.003$ & $0.44$ & $1.54 \pm 0.22$ & Gas Mass & (2)\\
Abell1758 & 0.279 & ... & (19) & 13:32:38.40 & +50:33:35.61 & $13.877 \pm 0.011$ & $0.486$ & $0.83 \pm 0.20$ & X-ray Luminosity & (1)\\
Abell1758 & 0.279 & ... & (19) & 13:32:52.04 & +50:31:34.62 & $14.178 \pm 0.016$ & $0.369$ & $0.83 \pm 0.20$ & X-ray Luminosity & (1)\\
MACS-J1347.5-1144 & 0.451 & ... & (1) & 13:47:29.38 & -11:45:06.0 & $14.449 \pm 0.010$ & $0.817$ & $3.16 \pm 0.44$ & Gas Mass & (2)\\
MACS-J1347.5-1144 & 0.451 & ... & (1) & 13:47:30.61 & -11:45:08.23 & $14.405 \pm 0.009$ & $0.851$ & $3.16 \pm 0.44$ & Gas Mass & (2)\\
Abell1835 & 0.253 & ... & (1) & 14:01:02.11 & +02:52:43.10 & $13.193 \pm 0.002$ & $0.769$ & $1.12 \pm 0.61$ & X-ray Temperature & (6)\\
MACS-J1427+44 & 0.487 & ... & (19) & 14:27:16.17 & +44:07:31.16 & $14.638 \pm 0.006$ & $0.769$ & ... & ... & ...\\
Abell1942 & 0.224 & ... & (7) & 14:38:21.88 & +03:40:13.10 & $13.178 \pm 0.003$ & $0.65$ & $0.34 \pm 0.19$ & X-ray Temperature & (6)\\
Abell1994 & 0.22 & ... & (12) & 14:56:13.47 & -05:48:55.80 & $13.939 \pm 0.005$ & $0.322$ & $0.60 \pm 0.15$ & X-ray Luminosity & (7)\\
MACS-J1652.3+5534 & ... & ... & ... & 16:52:18.67 & +55:34:58.54 & $14.547 \pm 0.009$ & ... & ... & ... & ...\\
RX-J1720.1+2638 & 0.164 & ... & (12) & 17:20:10.02 & +26:37:32.1 & $13.381 \pm 0.004$ & $0.35$ & $1.51 \pm 0.37$ & X-ray Luminosity & (9)\\
MACS-J1752.0+4440 & 0.366 & ... & (13) & 17:51:53.38 & +44:39:13.4 & $15.101 \pm 0.011$ & $0.299$ & ... & ... & ...\\
MACS-J1931.8-2634 & 0.352 & ... & (1) & 19:31:49.65 & -26:34:33.0 & $14.178 \pm 0.009$ & $0.599$ & $1.45 \pm 0.23$ & Gas Mass & (2)\\
ACT-CL-J2025.2+0030 & ... & 0.34 & (3) & 20:25:013.0 & +00:31:38.76 & $14.414 \pm 0.002$ & $0.481$ & $0.67 \pm 0.15$ & S-Z Parameter & (13)\\
ACT-CL-J2050.5-0055 & 0.622 & ... & (3) & 20:50:29.73 & -00:55:40.38 & $16.086 \pm 0.006$ & $0.339$ & $0.55 \pm 0.12$ & S-Z Parameter & (13)\\
ACT-CL-J2051.1+0056 & 0.333 & ... & (7) & 20:51:11.09 & +00:56:45.92 & $13.772 \pm 0.003$ & $0.742$ & $0.48 \pm 0.31$ & X-ray Luminosity & (3)\\
ACT-CL-J2055.4+0105 & 0.408 & ... & (7) & 20:55:23.23 & +01:06:7.830 & $14.925 \pm 0.002$ & $0.407$ & $0.98 \pm 0.52$ & X-ray Luminosity & (3)\\
ACT-CL-J2128.4+0135 & 0.385 & ... & (3) & 21:28:23.42 & +01:35:36.64 & $14.264 \pm 0.006$ & $0.647$ & $1.08 \pm 0.57$ & X-ray Luminosity & (3)\\
ACT-CL-J2129.6+0005 & 0.234 & ... & (7) & 21:29:39.95 & +00:05:21.32 & $13.203 \pm 0.004$ & $0.635$ & $1.23 \pm 0.47$ & X-ray Luminosity & (3)\\
ACT-CL-J2130.1+0045 & ... & 0.71 & (3) & 21:30:8.840 & +00:46:48.64 & $15.785 \pm 0.009$ & $0.541$ & $0.47 \pm 0.13$ & S-Z Parameter & (13)\\
MACS-J2140.1-2339 & 0.313 & ... & (3) & 21:40:15.03 & -23:39:37.8 & $13.842 \pm 0.003$ & $0.694$ & $0.69 \pm 0.09$ & Gas Mass & (2)\\
C1G-J214826.3-053312 & ... & 0.625 & (21) & 21:48:26.25 & -05:33:11.2 & $16.347 \pm 0.220$ & $0.266$ & $1.03 \pm 0.68$ & X-ray Luminosity & (12)\\
RX-J2149.3+0951 & ... & ... & ... & 21:49:19.68 & +09:51:36.79 & $15.008 \pm 0.010$ & ... & ... & ... & ...\\
ACT-CL-J2152.9-0114 & ... & 0.69 & (4) & 21:52:55.65 & -01:14:53.20 & $16.729 \pm 0.003$ & $0.226$ & $0.44 \pm 0.13$ & S-Z Parameter & (13)\\
Abell2390 & 0.233 & ... & (1) & 21:53:36.80 & +17:41:43.60 & $13.922 \pm 0.004$ & $0.327$ & $2.22 \pm 0.28$ & Gas Mass & (2)\\
Abell3827 & 0.0993 & ... & (7) & 22:01:53.44 & -59:56:42.60 & $12.323 \pm 0.001$ & $0.338$ & $0.68 \pm 0.17$ & X-ray Luminosity & (7)\\
MACS-J2211.7-0349 & 0.396 & ... & (7) & 22:11:45.88 & -03:49:44.3 & $14.143 \pm 0.008$ & $0.835$ & $2.64 \pm 0.36$ & Gas Mass & (2)\\
MACS-J2214.9-1359 & 0.5027 & ... & (7) & 22:14:57.25 & -14:00:12.8 & $14.938 \pm 0.011$ & $0.648$ & $1.51 \pm 0.82$ & X-ray Temperature & (4)\\
ACT-CL-J2220.7-0042 & ... & 0.57 & (4) & 22:20:47.05 & -00:41:54.11 & $15.713 \pm 0.008$ & $0.414$ & $0.36 \pm 0.14$ & S-Z Parameter & (13)\\
ACT-CL-J2229.2-0004 & ... & 0.61 & (4) & 22:29:7.540 & -00:04:10.50 & $15.701 \pm 0.002$ & $0.451$ & $0.39 \pm 0.14$ & S-Z Parameter & (13)\\
C1G-J223007.6-080949 & 0.623 & ... & (21) & 22:30:07.50 & -08:09:48.6 & $16.128 \pm 0.220$ & $0.282$ & $1.30 \pm 0.88$ & X-ray Luminosity & (12)\\
C1G-J223727.5+135523 & ... & 0.700 & (21) & 22:37:27.54 & +13:55:23.5 & $16.466 \pm 0.220$ & $0.239$ & $1.06 \pm 0.68$ & X-ray Luminosity & (12)\\
MACS-J2241.8+1732 & 0.3137 & ... & (23) & 22:41:56.32 & +17:32:6.980 & $14.079 \pm 0.002$ & $0.558$ & ... & ... & ...\\
CL2244 & 0.328 & ... & (23) & 22:47:13.38 & -02:05:41.40 & $14.836 \pm 0.009$ & $0.278$ & ... & ... & ...\\
MACS-J2248.7-4431 & 0.3475 & ... & (23) & 22:48:43.99 & -44:31:51.1 & $13.375 \pm 0.004$ & $1.253$ & ... & ... & ...\\
ACT-CL-J2253.3-0031 & ... & 0.54 & (4) & 22:53:24.26 & -00:30:30.71 & $15.862 \pm 0.035$ & $0.333$ & $0.39 \pm 0.13$ & S-Z Parameter & (13)\\
AC114 & 0.313 & ... & (4) & 22:58:56.33 & -34:45:40.4 & $14.022 \pm 0.003$ & $0.588$ & ... & ... & ...\\
ACT-CL-J2302.5+0002 & 0.52 & ... & (9) & 23:02:35.07 & +00:02:34.34 & $15.523 \pm 0.018$ & $0.417$ & $0.54 \pm 0.13$ & S-Z Parameter & (13)\\
Abell2552 & 0.302 & ... & (25) & 23:11:33.27 & +03:38:5.220 & $13.549 \pm 0.017$ & $0.775$ & $1.08 \pm 0.27$ & X-ray Luminosity & (1)\\
C1G-J231215.6+035307 & 0.648 & ... & (21) & 23:12:17.05 & +03:53:14.9 & $16.029 \pm 0.220$ & $0.357$ & $1.15 \pm 0.72$ & X-ray Luminosity & (12)\\
C1G-J231520.6+090711 & ... & 0.725 & (21) & 23:15:20.57 & +09:07:11.9 & $16.413 \pm 0.220$ & $0.266$ & $1.32 \pm 0.85$ & X-ray Luminosity & (12)\\
Abell2631 & 0.273 & ... & (12) & 23:37:39.73 & +00:16:16.90 & $14.237 \pm 0.006$ & $0.349$ & $0.98 \pm 0.24$ & X-ray Luminosity & (1)\\
ACT-CL-J2351.7+0009 & ... & 0.99 & (3) & 23:51:44.70 & +00:09:16.27 & $16.269 \pm 0.004$ & $0.623$ & $0.46 \pm 0.14$ & S-Z Parameter & (13)\\
		\hline
	\end{tabular}
\end{table}

\end{landscape}

\end{document}